\documentclass[conference]{IEEEtran}
\IEEEoverridecommandlockouts
\usepackage{amsmath,amssymb,amsfonts}
\usepackage{graphicx}
\usepackage{textcomp}
\usepackage{xcolor}
\usepackage{hyperref}
\usepackage{makecell}
\usepackage{float}  
\usepackage{marvosym}
\usepackage{microtype}
\usepackage{subcaption}
\usepackage{booktabs,tabularx,multirow,overpic}
\usepackage{svg}
\usepackage{adjustbox}
\usepackage{algorithm}
\usepackage{algpseudocode}
\algrenewcommand\algorithmiccomment[1]{\(\triangleright\) #1}
\usepackage{float}        
\usepackage[absolute,overlay]{textpos}
\usepackage[numbers,sort&compress]{natbib}
\usepackage{afterpage} 
\usepackage{titlesec}
\usepackage[utf8]{inputenc}
\DeclareUnicodeCharacter{202F}{\,}

\definecolor{color3}{rgb}{0.95,0.95,0.95}
\hypersetup{
    colorlinks=true,
    linkcolor=blue,  
    urlcolor=violet,  
    citecolor  = green,   
}
\def\BibTeX{{\rm B\kern-.05em{\sc i\kern-.025em b}\kern-.08em
    T\kern-.1667em\lower.7ex\hbox{E}\kern-.125emX}}

\newcommand{\StateLong}[1]{\State{\parbox[t]{\dimexpr\linewidth-\algorithmicindent}{#1\strut}}}

\begin{document}

\title{\huge{REACH: Reinforcement Learning for Efficient Allocation in Community and Heterogeneous Networks}}

\author{
    \IEEEauthorblockN{Zhiwei Yu$^1$, Chengze Du$^{1}$, Heng Xu$^1$, Ying Zhou$^2$, Bo Liu$^1$ , Jialong Li$^{1,}$\textsuperscript{\Letter}}
    \IEEEauthorblockA{$^1$ Computer Science and Control Engineering, Shenzhen University of Advanced Technology, Shenzhen, China \\
    \textsuperscript{2}School of Electronic Information Engineering, Beijing Jiaotong University\\
    }
    \IEEEauthorblockA{\text{\texttt{lijialong@suat-sz.edu.cn} 
    }}
}

\maketitle

\begin{abstract}
Community GPU platforms are emerging as a cost-effective and democratized alternative to centralized GPU clusters for AI workloads, aggregating idle consumer GPUs from globally distributed and heterogeneous environments. However, their extreme hardware/software diversity, volatile availability, and variable network conditions render traditional schedulers ineffective, leading to suboptimal task completion. In this work, we present REACH (Reinforcement Learning for Efficient Allocation in Community and Heterogeneous Networks), a Transformer-based reinforcement learning framework that redefines task scheduling as a sequence scoring problem to balance performance, reliability, cost, and network efficiency. By modeling both global GPU states and task requirements, REACH learns to adaptively co-locate computation with data, prioritize critical jobs, and mitigate the impact of unreliable resources. Extensive simulation results show that REACH improves task completion rates by up to 17\%, more than doubles the success rate for high-priority tasks, and reduces bandwidth penalties by over 80\% compared to state-of-the-art baselines. Stress tests further demonstrate its robustness to GPU churn and network congestion, while scalability experiments confirm its effectiveness in large-scale, high-contention scenarios.
\end{abstract}

\begin{IEEEkeywords}
Community GPU platforms; Reinforcement learning; Task scheduling; Distributed AI infrastructure
\end{IEEEkeywords}

\section{Introduction}
The advent of foundation models such as GPT-4~\cite{openai2023gpt4}, PaLM~\cite{Chowdhery2022,anil2023palm2}, and LLaMA~\cite{touvron2023llama,grattafiori2024llama3} has triggered an unprecedented demand for high-performance Graphics Processing Units (GPUs). Training these massive models is a monumental undertaking, with leading institutions routinely deploying over 10,000 A100 or H100 GPUs in a single run, incurring operational costs in the millions of dollars~\cite{Shen2025,Chong2025}. While public clouds like AWS~\cite{AWS_GPU_2025} offer on-demand access, the prohibitively high prices, ranging from 10 to 30 per GPU-hour, combined with global semiconductor supply constraints, have placed these critical resources beyond the reach of most smaller entities~\cite{datar2024promiseanalogdeeplearning}. This creates a significant barrier to entry for startups, academic research labs, and independent developers. The resulting concentration of compute power in the hands of a few well-funded institutions reinforces existing inequities and stifles the broader innovation ecosystem.

\begin{figure}[t]
  \centering
  \hspace*{-0.05\linewidth} 
  \noindent
  \begin{minipage}[t]{0.42\linewidth}  
    \centering
    \vspace{0pt}  
    \includegraphics[width=\linewidth]{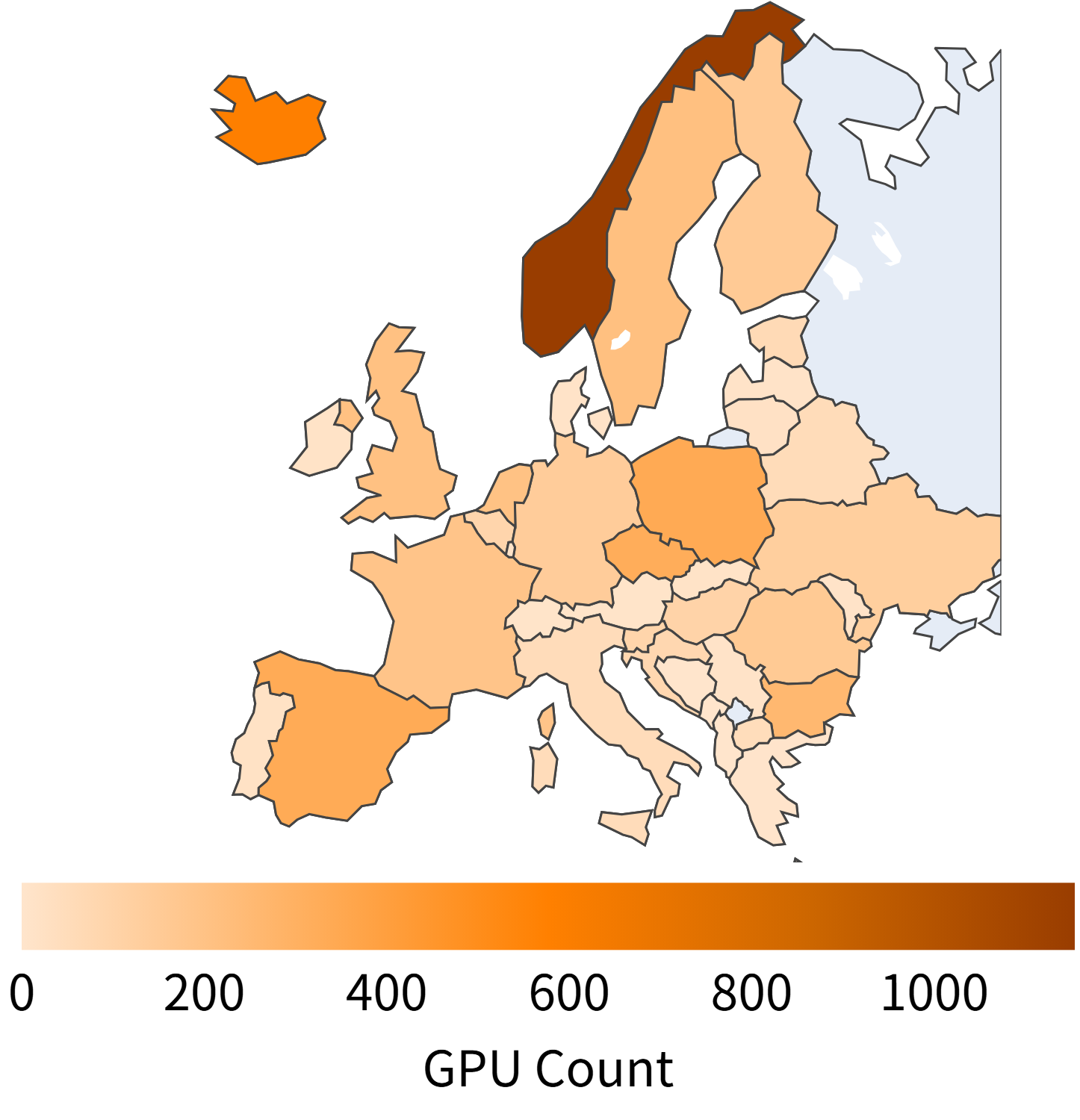}
    \makebox[\linewidth][l]{\hspace*{1.8em} \scriptsize \textbf{(a)} GPU Distribution in Europe}
  \end{minipage}%
  \hspace{0\linewidth}
  \begin{minipage}[t]{0.5\linewidth}  
    \centering
    \vspace{0pt}  
    \includegraphics[width=\linewidth]{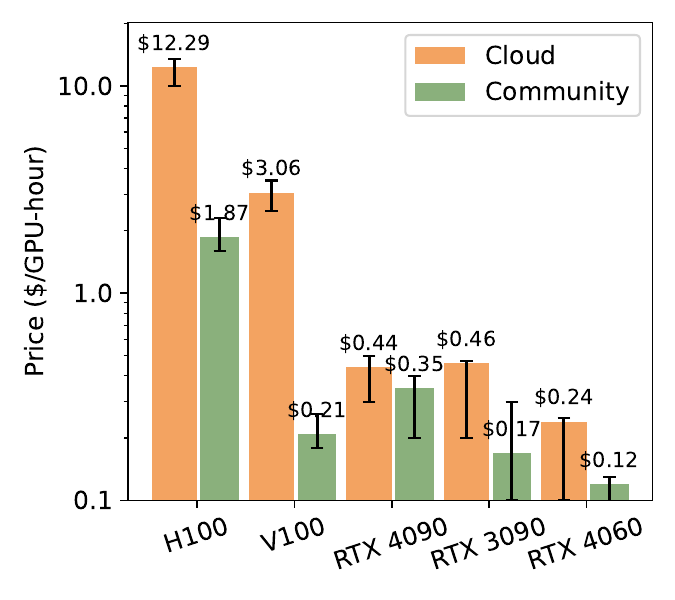}
    \raisebox{0.9mm}{\makebox[\linewidth][l]{\hspace*{1.8em} \scriptsize \textbf{(b)} Comparison of GPU Price}}

  \end{minipage}%
  \hspace{0\linewidth}
  \caption{Geographic distribution of (a) idle consumer GPUs in Europe and (b) price comparison between community and cloud offerings, showing up to 85\% savings for high-end models.}
  \label{fig:Distribution in China}
\end{figure}

Ironically, these expensive, centralized GPU clusters often suffer from chronic underutilization. Studies reveal that rigid gang scheduling, strict locality constraints, and inflexible resource allocation schemes lead to widespread fragmentation and idle capacity, even in compute-intensive AI scenarios~\cite{jeon2019multi,elvinger2025measuring}. This inefficiency highlights a fundamental disconnect: a vast, democratized demand for computation is being bottlenecked by a centralized, inefficiently managed supply.

In stark contrast to this scarcity, a massive, globally distributed resource remains largely untapped: the hundreds of millions of consumer-grade GPUs in personal computers, which are often idle for more than 20 hours a day~\cite{Brown2020}. This abundance of community GPUs, as shown in \autoref{fig:Distribution in China}, represents a vast potential in availability, distribution, and cost. This surplus has inspired platforms like Vast.ai~\cite{vastai2025} and SaladCloud~\cite{SaladCloud}, which aggregate idle community GPUs for general-purpose computing. These platforms have demonstrated the viability of this model, offering compute at prices from 60\% to 90\% below traditional cloud market rates. This presents a paradigm-shifting opportunity: to build a cost-effective, democratized, and scalable compute infrastructure for large-scale AI~\cite{gu2025deep,ismail2025survey,luo2023resource}.

However, harnessing this potential is a formidable task due to a unique set of challenges. At the foundational level, these networks are defined by extreme 
hardware and software heterogeneity, as community devices vary drastically in GPU model, memory capacity, driver versions, and operating systems. A scheduler must navigate this diversity to avoid bottlenecks. Compounding this, the unreliable availability of resources means they can disappear without warning due to host shutdowns or connectivity failures, making synchronous distributed training infeasible and necessitating robust fault tolerance mechanisms~\cite{Peng2024,Wei2024}. The network connecting these devices introduces a further layer of volatility, being subject to the inconsistent performance of residential ISPs, diurnal traffic patterns, and sudden congestion events. Consequently, traditional schedulers like Slurm or Kubernetes, which assume a stable, centralized environment, are rendered ineffective. They lack the necessary mechanisms for reliability prediction, dynamic policy adaptation, or incentive awareness in these volatile, decentralized networks.

To overcome these challenges, we propose \textbf{REACH (Reinforcement Learning for Efficient Allocation in Community and Heterogeneous Networks)}, an end-to-end deep reinforcement learning (RL) framework. REACH models the complex, multi-objective scheduling problem as a Markov Decision Process (MDP), capturing the stochastic nature of the environment. Critically, the state and action spaces of this MDP grow combinatorially and exponentially with the number of tasks and GPUs, making exact solutions computationally prohibitive for real-time scheduling. REACH circumvents this complexity by reframing the assignment task as a more tractable sequence scoring problem, reducing the action space from combinatorial ($\mathcal{O}(C_{N}^{k})$) to linear ($\mathcal{O}(N)$). It leverages a Transformer-enhanced Actor-Critic architecture to learn a sophisticated scheduling policy that balances performance, cost, reliability, and network topology in real time.

\begin{figure}[tbp]
    \centering
    \includegraphics[width=9cm]{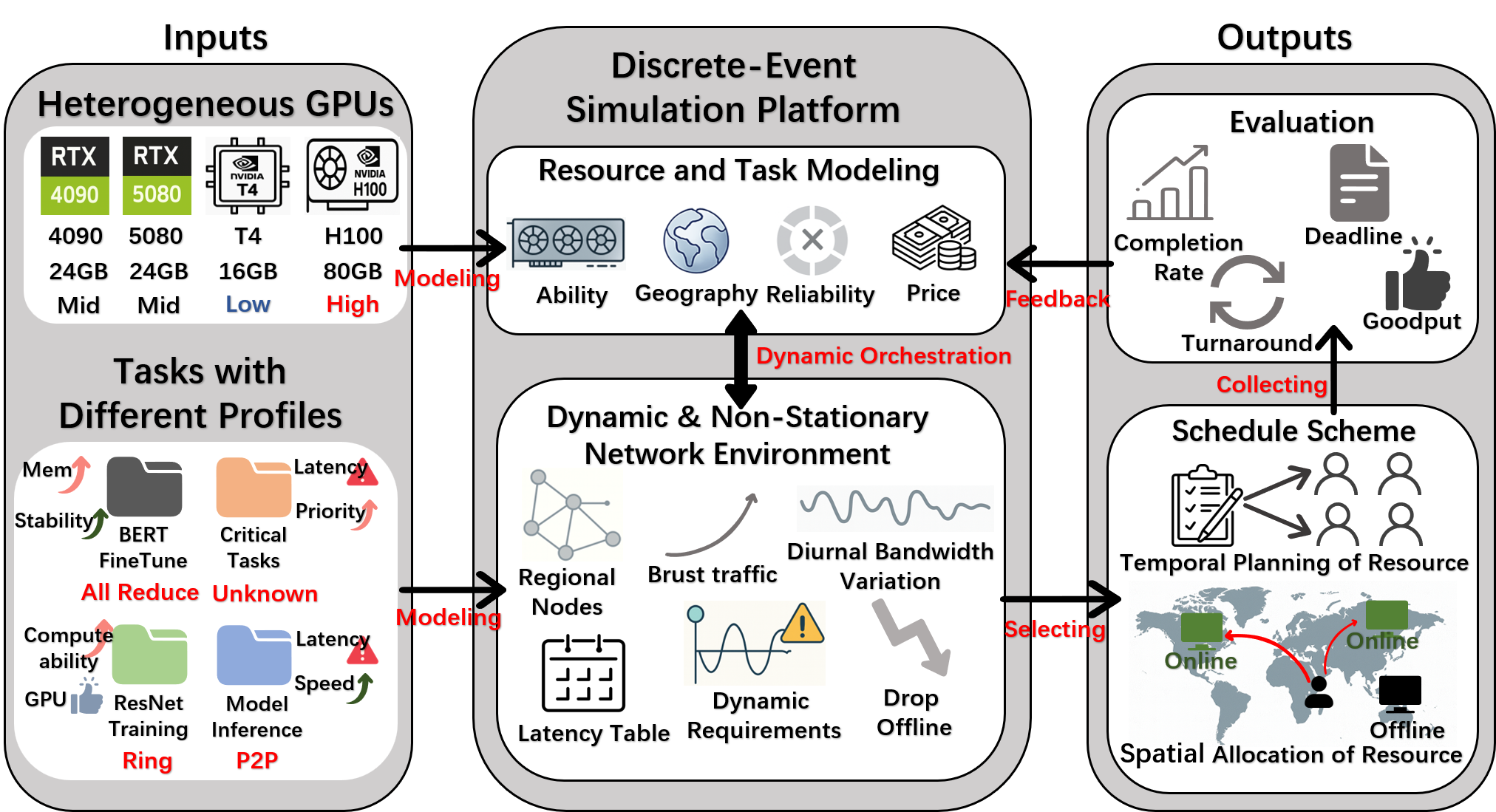}
    \caption{
    Dynamic GPU scheduling framework combining heterogeneous resources, network-aware simulation, and QoS-driven decision-making.}
    \label{fig:Figure 5}
\end{figure}

We conducted a rigorous evaluation of REACH using a comprehensive, discrete-event simulation platform designed to model the dynamics of community GPU networks. As illustrated in \autoref{fig:Figure 5}, this platform integrates heterogeneous GPU and task modeling with a dynamic, non-stationary network environment, enabling realistic simulation of compute performance, network latency/bandwidth fluctuations, and node reliability. It supports dynamic orchestration of tasks under diverse QoS objectives, capturing how scheduling decisions impact key metrics such as completion rate, deadline satisfaction, and GoodPut. This design bridges realistic system modeling with algorithmic decision-making, providing a reproducible and controllable testbed for training and benchmarking REACH under complex, real-world–like conditions.

Our experiments were designed to assess efficiency, task-awareness, and robustness. The results demonstrate that REACH establishes a new state-of-the-art in scheduling for heterogeneous, unreliable environments. Our key contributions are summarized as follows:

\begin{enumerate}
    \item \textbf{A Novel RL-based Scheduler:} We designed and implemented REACH, a Transformer-based agent that learns to make intelligent, context-aware scheduling decisions, significantly outperforming traditional algorithms.
    \item \textbf{State-of-the-Art Performance:} Under equivalent loads, REACH \textbf{improves the overall task completion rate by over 17\%} compared to baselines. For high-priority critical tasks, it more than doubles the performance, \textbf{boosting the completion rate from 30.3\% to 63.6\%}.
    \item \textbf{Unprecedented Robustness:} REACH shows exceptional resilience. In an extreme stress test where the \textbf{GPU dropout rate was increased by 16$\times$}, the performance of baseline schedulers collapsed. In contrast, REACH maintained a \textbf{over 95\% deadline satisfaction rate}, proving its ability to guarantee Quality of Service (QoS) in volatile conditions.
    \item \textbf{Topology-Aware Cost Reduction:} The agent intelligently co-locates computation with data, allowing over 80\% of its scheduled tasks to avoid the high bandwidth penalties that plague other strategies.
\end{enumerate}

These compelling quantitative results prove that REACH can effectively tame the inherent uncertainty of decentralized resources. It paves the way for community computing to become a viable, reliable, and cost-effective alternative to centralized clouds, taking a significant step towards the democratization of large-scale AI.

\section{Related Work}

Our work is situated within the field of GPU scheduling for deep learning, particularly focusing on distributed and heterogeneous environments. We review the literature from two primary perspectives: scheduling in traditional centralized data centers, and resource scheduling in emerging community and decentralized networks.

\subsection{Scheduling in Centralized GPU Clusters}
In centralized data centers, GPU scheduling has been extensively studied, with numerous works analyzing multi-tenant cluster workloads and revealing inefficiencies from rigid gang scheduling, strict locality constraints, and co-location interference~\cite{elvinger2025measuring,weng2023fragmentation,du2025crosstimeslotoptimizationdistributedgpu}. Such factors cause severe resource fragmentation and underutilization, often below 50\%, with fragmentation delays alone contributing up to 80\% of total queuing time~\cite{jeon2019multi}. While hardware-level sharing mechanisms like NVIDIA MIG alleviate internal fragmentation, they exacerbate external fragmentation, motivating packing algorithms that explicitly optimize fragmentation state~\cite{weng2023fragmentation,weng2023fragmentation_mdpi}.

To improve utilization, prior research has pushed scheduling granularity toward the hardware, exploring iteration-, operator-, and kernel-level sharing, as well as elastic GPU allocation~\cite{yu2020salus,strati2024orion,coppock2025lithos,cheng2021elastic}. Another line addresses heterogeneous, multi-objective scheduling, balancing performance, fairness, and topology awareness~\cite{narayanan2020gavel,sultana2024hadar,li2022astraea,yang2025gpu,mo2025fast,mahajan2020themis,chaudhary2020gandivafair,hwang2023ark,rajasekaran2024cassini,garcia2018topology,ryu2020topology,pan2022efficient,li2024optimizing}. Solutions range from economic mechanisms to topology-aware placement, and more recently, machine learning-based approaches using GNNs and reinforcement learning~\cite{zhang2023tag,gu2025deep,ryu2020topology}. However, these schedulers assume stable, high-performance infrastructure with predictable resource availability~\cite{zheng2022survey,slurm_pain_points}, making them ill-suited for the volatility and unreliability of decentralized community networks.

\begin{figure}[t]
  \centering
  \scalebox{0.8}{\includegraphics[width=\linewidth]{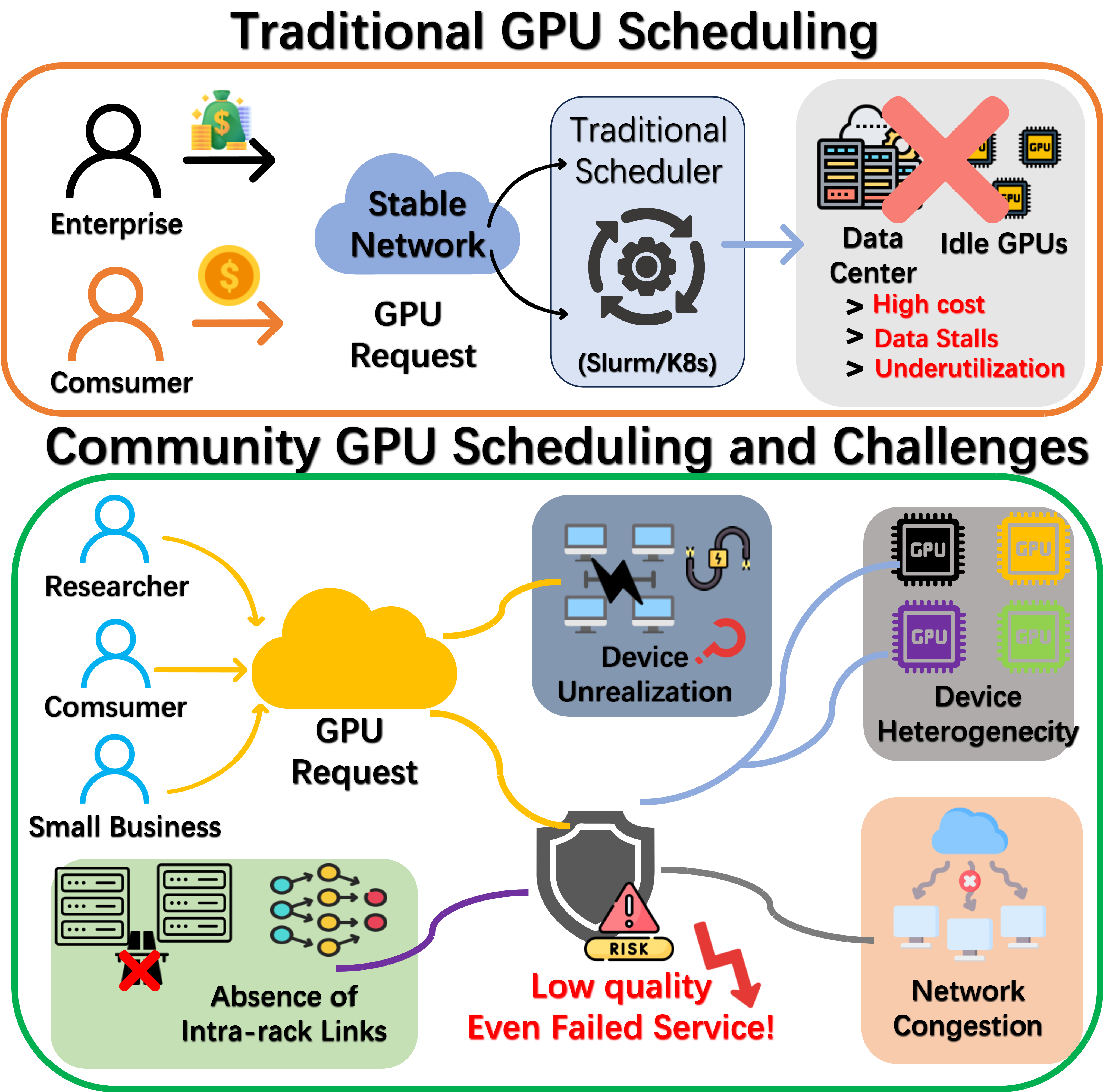}}
  \caption{
    Traditional scheduling suffers from high costs and underutilization; community GPU scheduling faces reliability and heterogeneity challenges impacting service quality.
  }
  \label{fig:Tradition GPU scheduling and communtity GPU}
\end{figure}

\subsection{Resource Scheduling in Community Networks}
Parallel to centralized systems, volunteer and community computing harnesses idle consumer-grade GPUs worldwide~\cite{anderson2020volunteer,atre2021distributed,lhanafi2016volunteer}, with roots in projects like \textbf{SETI@home} that demonstrated large-scale distributed analysis~\cite{anderson2002setihome}. The Berkeley Open Infrastructure for Network Computing (BOINC) generalized this model for hundreds of scientific projects~\cite{korpela2012boinc}. Recently, commercial platforms such as Vast.ai and SaladCloud aggregate idle GPUs for general-purpose computing at significantly lower costs than traditional clouds, underscoring the economic potential of this globally distributed resource pool~\cite{vastai2025,SaladCloud,Bains2025,Chong2025}.

While resource aggregation has been successful, scheduling in community networks remains relatively simple, often relying on static pricing and manual matching. This is largely due to two challenges uncommon in centralized clusters: \textbf{extreme heterogeneity} in hardware and software configurations, which complicates compatibility and performance predictability~\cite{slurm_pain_points}, and \textbf{unreliable availability} (``churn''), where non-dedicated resources can disappear unexpectedly due to host or network failures~\cite{Peng2024}. Such volatility severely impacts long-running distributed tasks, and early works recognized the inadequacy of naive First-Come-First-Serve policies, proposing reactive, threshold-based reliability filters~\cite{estrada2006scheduling}. However, modern deep learning workloads demand more sophisticated, probabilistic models that weigh stability against performance, enabling choices like preferring slightly slower but more reliable nodes.

These two primary challenges—extreme device heterogeneity and unreliable availability—fundamentally differentiate community GPU networks from their centralized counterparts. As summarized in \autoref{fig:Tradition GPU scheduling and communtity GPU}, in addition to these, community platforms often suffer from the absence of high-bandwidth intra-rack links and frequent network congestion, both of which can further exacerbate latency and reduce effective throughput. 

In summary, while sophisticated schedulers are well-established in centralized computing, they are ill-suited to the volatile and unreliable nature of community networks. Conversely, existing community computing approaches embrace these dynamic environments but lack the automated, reliability-aware scheduling necessary for mission-critical workloads. As comprehensive surveys confirm, most solutions still target centralized clusters~\cite{zheng2022survey,liang2024resource}. Our work addresses this gap by designing an intelligent scheduler for robust and efficient decision-making in these challenging networks.

\section{REACH algorithm}
To address this, we first formulate the scheduling problem as a MDP, capturing the stochastic, multi-objective nature of task placement in community GPU networks.
Building on this formulation, we present REACH, a reinforcement learning algorithm that leverages a Transformer-enhanced Actor-Critic architecture to make efficient, adaptive placement decisions for our tasks in real time.


\subsection{Problem Formulation}


Let the community GPU platform be defined by a set of heterogeneous GPUs $\mathcal{G}$, a dynamic queue of incoming tasks $\mathcal{T}$, and an unreliable network backbone $\mathcal{N}$. More precisely, the community platform is composed of:

\textbf{A Heterogeneous GPU Set ($\mathcal{G}$).} 
A set of $N$ GPU resources where each resource $g_i \in \mathcal{G}$ is characterized by a comprehensive attribute tuple: $g_i = (C_i, M_i, L_i, P_i, \delta_i(t))$. These attributes explicitly capture the system's heterogeneity: $C_i$ is the raw compute power; $M_i$ is the available memory, which constrains the maximum model size and batch capacity; $L_i$ is the geographical location, which determines data gravity and network latency; $P_i$ is the economic cost model; and $\delta_i(t)$ represents the dynamic reliability, or dropout probability, modeling the unpredictable nature of community resources that can vanish without warning.
 
\textbf{A Diverse Task Set ($\mathcal{T}$).} 
A dynamic queue of $M$ incoming tasks. Each task $T_j \in \mathcal{T}$ is defined by its requirements: $T_j = (R_j, M_j^{\text{req}}, D_j, K_j, \Omega_j, L_j^{\text{data}})$. Here, $R_j$ and $M_j^{\text{req}}$ specify the scale of required resources. The absolute deadline $D_j$ and criticality flag $K_j$ define the task's urgency and business impact. The communication profile $\Omega_j$ and data location $L_j^{\text{data}}$ are critical for network-aware scheduling, as ignoring them can lead to suboptimal placement and degraded performance, particularly for tasks with strong communication dependencies.
    
\textbf{An Unreliable Network Backbone ($\mathcal{N}$).} 
Modeled as a graph where inter-regional links are defined by heterogeneous, geography-dependent latency and bandwidth. The network state is non-stationary, capturing the inconsistent performance of residential ISPs by incorporating systematic diurnal traffic patterns, short-term stochastic fluctuations, and sudden, high-impact congestion events on random links.

The central challenge is to design a scheduling policy \textbf{$\pi$} that maps tasks to resources over time to maximize a multi-objective utility function.

At any discrete decision epoch $t$, the scheduler observes the system state $s_t \in S$. This state encapsulates the complete status of the platform, including the attributes and availability of all GPUs, the requirements of all pending tasks, and the current condition of the network. The scheduler then takes an action $a_t \in \mathcal{A}$, which consists of assigning a subset of pending tasks to available GPUs. This action causes the system to transition to a new state $s_{t+1}$ and yields a reward $R_{t+1}$.

\begin{figure}[t]
  \centering
  \hspace*{-0.28\linewidth} 
  \noindent
  \begin{minipage}[t]{0.42\linewidth}  
    \vspace{0pt}  
    \makebox[\linewidth][l]{\scriptsize \textbf{(a)} MDP Complexity Explosion}
    \includegraphics[width=\linewidth]{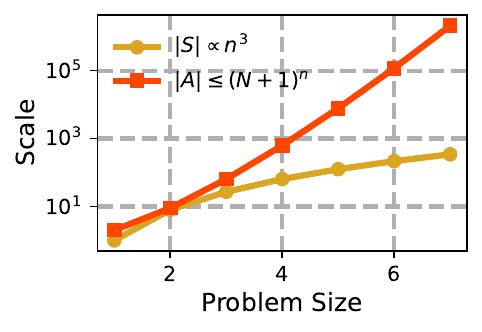}
  \end{minipage}%
  \hspace{0\linewidth}
  \begin{minipage}[t]{0.4\linewidth}
    \centering
    \vspace{0pt}
    \makebox[\linewidth][l]{\scriptsize \textbf{(b)} Configuration Parameters}
    \vspace{0.3em}
    \scriptsize
    \begin{tabular}{@{}p{1.0cm}p{1.9cm}p{1.6cm}@{}}
      \toprule
      \textbf{Component} & \textbf{Description} & \textbf{Scale / Type} \\
      \midrule
      GPU & Compute, memory & $N$ GPUs \\
      Tasks & Demand, deadlines & $M$ Tasks \\
      Network & Latency, bandwidth & Graph \\
      Reliability & Availability, power & $\delta_i(t) \in [0,1]$ \\
      Others & Load, congestion & 3--5D vector \\
      \bottomrule
    \end{tabular}
  \end{minipage}
  \caption{(a) The MDP scheduling problem becomes intractable at scale. 
  (b) Key parameters required to model the community GPU scheduling problem.}
  \label{fig:mdp-complexity-configuration}
\end{figure}

This sequential decision-making process can be formally modeled as a high-dimensional, stochastic MDP. The objective is to find an optimal policy $\mathit{\pi^*} : \mathcal{S} \longrightarrow  \mathcal{A}$  that maximizes the expected discounted cumulative reward over a infinite horizon:
\begin{equation}
\pi^* = \arg\max_{\pi} \mathbb{E}_{\pi} \left[ \sum_{t=0}^{\infty} \gamma^t R(s_t, a_t) \right]
\end{equation}
where $\gamma \in \left [ 0,1 \right ]$ is the discount factor. To optimize our multi-faceted goals, we define a reward function \( R(s_t, a_t) \) that reflects the platform’s multi-dimensional goals. As established, simple metrics like task completion count are insufficient. Instead, the reward is defined as a weighted sum of multiple performance indicators, reflecting the true utility generated by the platform:
\begin{equation}
\label{eq:reward_function} 
\begin{aligned}
  R(s_t,& a_t) = w_{comp} \cdot (I_{ontime} + I_{late}) + w_{deadline} \cdot I_{ontime} \\
                   & + w_{fail} \cdot I_{fail} + w_{cost} \cdot C_{norm} 
                    + w_{comm} \cdot (P_{comm} - 1)
\end{aligned}
\end{equation}
Here, the $w$ terms are tunable weights for each component. $I_{ontime}$, $I_{late}$, and $I_{fail}$ are indicator functions that are triggered based on the task's final status (e.g., successful on-time, successful but late, or failed). $C_{norm}$ represents the normalized operational cost, while $P_{comm}$ is a communication penalty factor ($P_{comm} \ge 1$) that penalizes assignments to geographically dispersed GPUs. In training, $R(s_t, a_t)$ serves as the immediate reward at each time step.
These rewards are accumulated into the return $\hat{R}_t$, which is then used to compute the advantage
$\hat{A}_t = \hat{R}_t - V(s_t)$ for policy optimization. This formulation ensures the agent is optimized not just for task completion, but for a balance of timeliness, cost-efficiency, and network-aware allocation.

Although the underlying MDP can, in theory, be solved using exact methods, doing so is computationally prohibitive in practice. As illustrated in~\autoref{fig:mdp-complexity-configuration}, The state space $\mathcal{S}$ grows combinatorially with the number of tasks and GPUs, while the action space $\mathcal{A}$—which represents all valid task-to-GPU assignments—expands exponentially. While tractable solutions may exist offline or in small-scale settings, such exhaustive computation is unsuitable for real-time scheduling scenarios. In latency-sensitive environments like networked GPU clusters, such delays would directly affect user experience. This necessitates a more efficient and adaptive approach, motivating our proposed algorithm.

\begin{figure}[tbp]
    \centering
    \includegraphics[width=9cm]{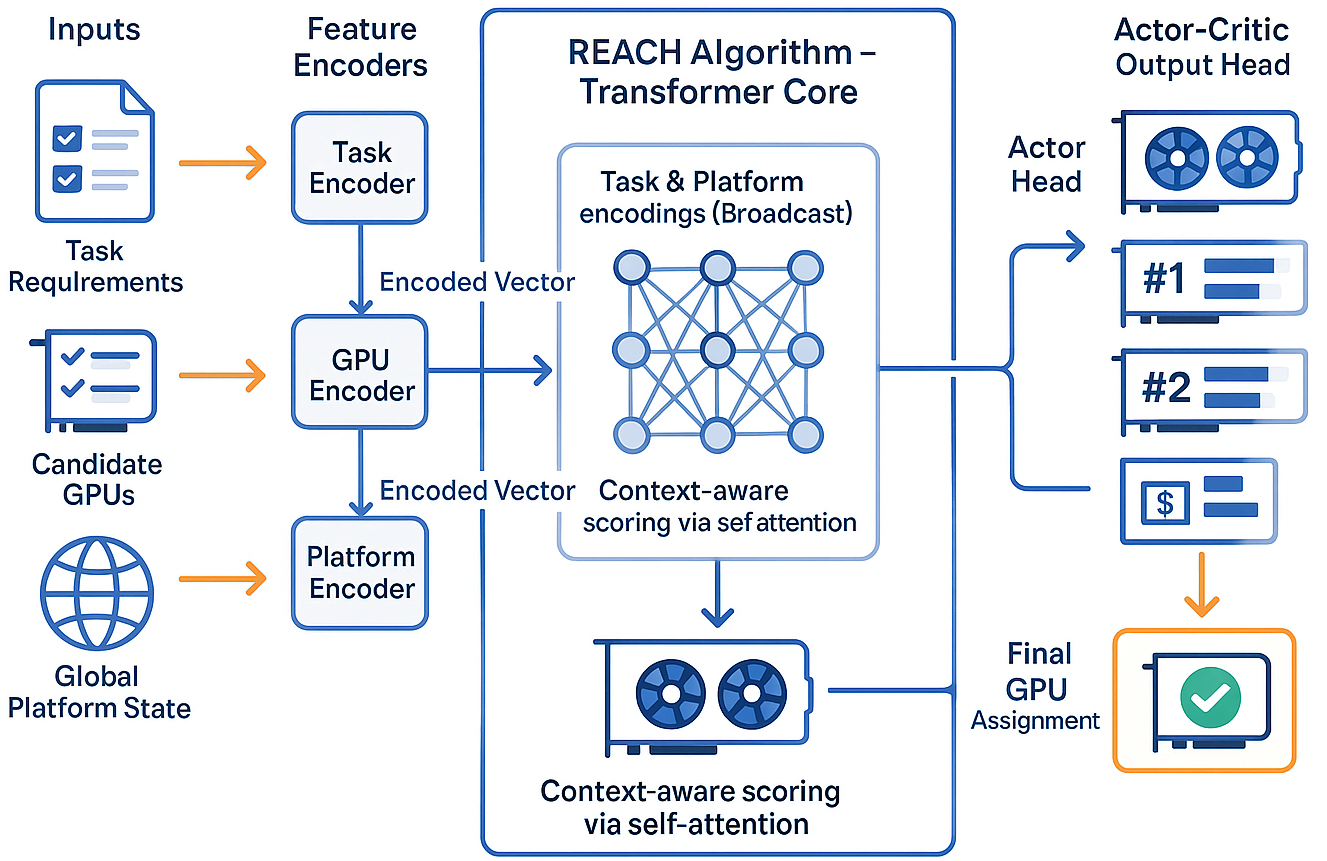}
    \caption{REACH’s Transformer-based Actor–Critic ranks GPUs and evaluates state value for optimal scheduling.}
    \label{fig:Figure 4}
\end{figure}


\subsection{Transformer-based Actor-Critic Architecture}
To address the aforementioned challenges, we propose REACH, an end-to-end deep reinforcement learning scheduling algorithm. The core idea of REACH is to \textbf{reframe the complex combinatorial assignment problem into a more tractable sequence scoring problem}. For each task to be scheduled, instead of searching directly within the vast GPU combination space, REACH scores all candidate GPUs that meet the basic requirements and selects the top-$k$ highest-scoring GPUs as the final allocation. 
\begin{equation}
    a_t = \mathrm{Top}\text{-}k \left( \pi_\theta(s_t) \right)
\end{equation}
where $k$ is the number of GPUs required by the task, and $\pi_\theta$ outputs a score for each candidate GPU. 
This approach reduces the action space from a combinatorial level, $\mathcal{O}(C_N^k)$, to a linear level, $\mathcal{O}(N)$, enhancing decision-making efficiency and scalability. To enable intelligent scoring, REACH employs a Transformer-based Actor-Critic neural network architecture, as illustrated in \autoref{fig:Figure 4}.

Beyond its empirical performance, the adoption of a Transformer core is grounded in theoretical considerations. In particular, its self-attention mechanism can model all-pair dependencies between candidate GPUs in a single layer. This global receptive field allows the policy to account for complex interactions—such as topology proximity, bandwidth constraints, and reliability correlations—that are hard to capture with local MLP structures. In heterogeneous, geographically distributed GPU networks, such long-range dependency modeling is essential for optimal scheduling.

\noindent \textbf{Input Encoders.} REACH first vectorizes heterogeneous state information through a dedicated feature engineering process. Continuous numerical attributes (e.g., memory requirements, deadlines) are normalized to a consistent range. Categorical data, such as geographical location and task communication topology, is converted into one-hot vectors to avoid imposing a false ordinal relationship. Crucially, to capture the dynamic reliability of resources, the model incorporates temporal features like time ``since offline'' and ``online duration''. These engineered feature vectors are then fed into separate linear layers (MLPs) to be encoded into high-dimensional embeddings for the Transformer core.

\noindent \textbf{Transformer Encoder.} The embeddings of all candidate GPUs are combined with the broadcasted task and global context embeddings through element-wise addition, forming a sequence that is input to the Transformer encoder. The central component of this architecture is the self-attention mechanism. This mechanism enables the model to perform a holistic evaluation, allowing the representation of each candidate GPU to be contextually informed by its relationship with all other available GPUs. For instance, the model can learn that assigning a task to geographically dispersed GPUs will incur high communication costs, even if each individual GPU is excellent.

Formally, let $f^{task}$, $f^{gpu}_i$, and $f^{global}$ denote the task, GPU, and global context feature vectors respectively. 
We project them to a shared $d_{\mathrm{model}}$-dimensional space and combine them:
\begin{equation}
    h_i^{(0)} = W_g f^{gpu}_i + W_t f^{task} + W_c f^{global}, \quad i=1,\dots,N
\end{equation}
The Transformer encoder then applies multi-head self-attention:
\begin{equation}
    \mathrm{MHA}(H) = \mathrm{Concat}(\mathrm{head}_1,\dots,\mathrm{head}_h)W^O
\end{equation}
\begin{equation}
    \mathrm{head}_j = \mathrm{Attention}(HW_j^Q, HW_j^K, HW_j^V)
\end{equation}
producing embeddings $h_i^{(L)}$ for all $N$ candidate GPUs.

\noindent \textbf{Output Heads.} After processing by the Transformer, the contextualized GPU embeddings $\{h_i^{(L)}\}_{i=1}^N$ are fed into two independent output heads:

\begin{itemize}
    \item \textbf{Actor Head}: A linear projection outputs a scalar logit for each GPU:
    \begin{equation}
        z_i = W_a h_i^{(L)}, \quad i=1,\dots,N
    \end{equation}
    The logits form a probability distribution over candidate GPUs via the softmax function:
    \begin{equation}
    \pi_\theta(a_t|s_t) = \mathrm{softmax}(z_1,\dots,z_N)
    \end{equation}
    where $\pi_\theta(a_t|s_t)$ is the policy, i.e., the probability of selecting each GPU given state $s_t$.

    \item \textbf{Critic Head}: An aggregated representation of the current state is obtained by averaging all contextualized GPU embeddings:
    \begin{equation}
    \bar{h}^{(L)} = \frac{1}{N} \sum_{i=1}^N h_i^{(L)}
    \end{equation}
    The critic outputs the estimated state value:
    \begin{equation}
    V_\phi(s_t) = W_v \bar{h}^{(L)}
    \end{equation}
\end{itemize}

\subsection{Learning Algorithm}

REACH is trained using the \textbf{Proximal Policy Optimization (PPO)} algorithm, an advanced Actor-Critic method known for its training stability and sample efficiency. In each training step, REACH collects a batch of experiences (containing states, actions, rewards, etc.) from its interaction with the environment. The immediate reward at time $t$ is denoted as $R(s_t, a_t)$, and the empirical return is computed as:
\begin{equation}
\hat{R}_t = \sum{l=0}^{T-t} \gamma^l R(s_{t+l}, a_{t+l}),
\label{eq:return}
\end{equation}
where $\gamma \in (0,1]$ is the discount factor. The (unnormalized) advantage estimate is:
\begin{equation}
\hat{A}_t = \hat{R}_t - V\phi(s_t).
\label{eq:advantage}
\end{equation}
Before being used in the policy update, advantages are normalized within each mini-batch to stabilize training:
\begin{equation}
\hat{A}_t^{\text{norm}} = \frac{\hat{A}_t - \mu_A}{\sigma_A + \epsilon},
\label{eq:advantage_norm}
\end{equation}
where $\mu_A$ and $\sigma_A$ denote the mean and standard deviation of $\hat{A}_t$ over the batch, and $\epsilon$ is a small constant for numerical stability.

PPO updates the Actor network parameters using the clipped surrogate objective:
\begin{align}
L^{\text{PPO}}(\theta) 
&= \mathbb{E}_t \bigg[
    \min \Big(
        r_t(\theta) \, \hat{A}_t^{\text{norm}}, \nonumber\\
&\qquad\quad
        \text{clip}\!\big(r_t(\theta),\, 1-\epsilon,\, 1+\epsilon\big) 
        \, \hat{A}_t^{\text{norm}}
    \Big)
\bigg],
\label{eq:ppo_objective}
\end{align}
where the importance sampling ratio is
\begin{equation}
r_t(\theta) = \frac{\pi\theta(a_t \mid s_t)}{\pi{\theta{\text{old}}}(a_t \mid s_t)}.
\label{eq:importance_ratio}
\end{equation}
The critic network is optimized via the value loss:
\begin{equation}
L^{\text{value}}(\phi) = \mathbb{E}_t \left[ \big( V\phi(s_t) - \hat{R}_t \big)^2 \right],
\label{eq:value_loss}
\end{equation}
and an entropy bonus is included to encourage exploration.
\begin{equation}
L^{\text{entropy}}(\theta) = \mathbb{E}_t \left[ \mathcal{H}\big(\pi\theta(\cdot \mid s_t)\big) \right]
\label{eq:entropy_bonus}
\end{equation}

The final total loss is:
\begin{equation}
L_{\text{total}}(\theta, \phi) = -L^{\text{PPO}}\big(\theta, \hat{A}_t^{\text{norm}}\big) + c_v L^{\text{value}}(\phi) - c_e L^{\text{entropy}}(\theta),
\label{eq:total_loss}
\end{equation}
where $c_v$ and $c_e$ are coefficients that balance the value and entropy terms.

\begin{algorithm}[t]
\caption{REACH Algorithm}
\label{alg:REACH_scheduler_compressed}
\begin{algorithmic}[1]
\Require Simulation environment $\mathcal{P}$, Total training steps $T_{\text{total}}$
\State Initialize Actor-Critic $(\pi_\theta, V_\phi)$, optimizer, replay buffer $\mathcal{B}$, pending decisions $D_{\text{pending}}$.

\For{each training step $t=1, \dots, T_{\text{total}}$}
    \State Wait for a task $T_j$ and identify candidate GPUs $G_{\text{cand}}$.
    \If{$|G_{\text{cand}}| < T_j.\text{gpu\_required}$} \textbf{continue} \EndIf
    
    \StateLong{Sample action $a_t \sim \pi_\theta(\cdot|s_t)$, store context $(s_t, a_t, \dots)$ in $D_{\text{pending}}$, and dispatch $T_j$.}
    
    \If{an outcome for task $T_k$ is received}
        \State Retrieve context, calculate reward $r_k$, add to $\mathcal{B}$.
    \EndIf
    
    \If{$|\mathcal{B}| \ge \text{BATCH\_SIZE}$}
        \State Sample a mini-batch from $\mathcal{B}$. 
        \State Calculate advantages $\hat{A}_i$.
        \For{epoch $= 1, \dots, \text{PPO\_EPOCHS}$}
            \State Compute the PPO loss $L$ on the mini-batch.
            \State Update parameters $(\theta, \phi)$ using the loss $L$.
        \EndFor
        \State Clear the replay buffer $\mathcal{B}$.
    \EndIf
\EndFor
\end{algorithmic}
\end{algorithm}

During training, once the replay buffer $\mathcal{B}$ is filled with a batch of experiences, PPO performs $K$ update epochs. In each epoch, the batch is divided into mini-batches, and the parameters are updated by minimizing \autoref{eq:total_loss}. This repeated mini-batch optimization improves sample efficiency while the clipping term in \autoref{eq:ppo_objective} prevents destructive policy oscillations.

Our complete training and inference pipeline is summarized in ~\autoref{alg:REACH_scheduler_compressed}. The process operates in an event-driven loop. Upon the arrival of a new task $T_j$, the system first identifies a set of candidate GPUs, $G_{\text{cand}}$, that meet its basic requirements. Subsequently, the Actor network, $\pi_\theta$, samples an action $a_t$ (i.e., a specific GPU assignment) based on the current system state $s_t$ and dispatches the task to the environment. To handle the asynchronous feedback, this decision context is temporarily stored in a pending queue, $D_{\text{pending}}$.

\section{Platform Design}

To rigorously evaluate scheduling algorithms in a reproducible manner, we have developed a comprehensive simulation platform for community GPU computing. Architected as a discrete-event system, it captures the complex dynamics of such ecosystems by abstracting key entities—like heterogeneous GPUs and diverse workloads—and simulating emergent properties, such as non-stationary network conditions and stochastic resource availability. This provides a robust testbed for assessing scheduler performance under realistic and challenging conditions.

\begin{table}[t]
\caption{Representative GPU Models and Characteristics}
\label{tab:gpu_specifications_final}
\centering
\small
\scalebox{0.9}{
\begin{tabular}{lcccc}
\toprule
 \textbf{GPU} & \textbf{Memory} & \textbf{Tensor32} & \textbf{Available} & \textbf{Est. Hourly} \\
  \textbf{Type} & \textbf{(GB)} & \textbf{(TFLOPS)} & \textbf{Quantity} & \textbf{Cost (USD)} \\
\midrule
H100     & 80 & 989.0 & 45   & $\sim$\$2.26 \\
RTX 4090 & 24 & 82.6 & 2064 & $\sim$\$0.40 \\
RTX 3080 & 12 & 29.8 & 128  & $\sim$\$0.09 \\
RTX 3060 & 12 & 12.4 & 654  & $\sim$\$0.06 \\
\bottomrule
\end{tabular}
}
\end{table}

\subsection{Instantiating MDP Elements: From Theory to Models}
Our platform's primary goal is to translate the abstract components of our MDP into concrete, simulated entities. This ensures that the challenges our agent faces are not merely theoretical but are grounded in realistic operational complexities that demand intelligent, adaptive solutions.


\noindent \textbf{GPU as a Techno-Economic Asset.}
The GPU attribute tuple defined in our MDP, $g_{i}=(C_{i},M_{i},L_{i},P_{i},\delta_{i}(t))$, is modeled as a multi-faceted techno-economic asset. This approach moves beyond simple hardware specifications to capture the true trade-offs in a community market.~\autoref{tab:gpu_specifications_final} lists the representative GPU models that constitute this heterogeneous resource pool. 

The \textbf{theoretical location} ($L_i$) and \textbf{economic cost} ($P_i$) are instantiated with specific geographical coordinates and an associated cost model, including data egress costs per gigabyte and a base hourly rate. This design creates the tangible performance-versus-cost trade-off that the agent must learn to navigate, directly corresponding to the $w_{\text{cost}} \cdot C_{\text{norm}}$ term in our reward function. 

The \textbf{dynamic reliability} ($\delta_i(t)$) is implemented as a stochastic dropout probability for each GPU. This directly simulates the Unreliable Availability challenge, testing the agent's ability to develop robust policies that mitigate the impact of the failure penalty, $w_{\text{fail}} \cdot I_{\text{fail}}$, in the reward function.

\newcolumntype{C}[1]{>{\centering\arraybackslash}p{#1}}

\begin{table}[t]
\caption{Representative Workload Examples}
\label{tab:task_workloads}
\centering
\small
\scalebox{0.9}{
\begin{tabular}{lccl}
\toprule
\textbf{Workload Name} & \textbf{Base Time} & \textbf{GPUs} & \textbf{Comm. Profile} \\ 
\midrule
Critical Inference & 0.1 h & 1  & Point-to-Point  \\
BERT Finetune    & 6.0 h  & 1  & Compute-Heavy \\
LLaMA-7B Finetune & $>$12.0 h  & 16  & All-Reduce \\
ResNet Training   & $>$12.0 h  & 32 & Ring (High) \\
\bottomrule
\end{tabular}
}
\end{table}


\noindent \textbf{Tasks with Embedded Communication Profiles.}
Similarly, each computational request is a practical implementation of the theoretical task tuple $T_{j}=(R_{j},M_{j}^{\text{req}},D_{j},K_{j},\Omega_{j},L_{j}^{\text{data}})$.~\autoref{tab:task_workloads} shows a sample of these diverse workloads and their corresponding profiles.

The task's \textbf{deadline} ($D_j$) and \textbf{criticality flag} ($K_j$) are modeled as explicit attributes, allowing us to generate more realistic workloads with diverse Service Level Objectives. This is crucial for evaluating the agent's ability to optimize for timeliness, as directly incentivized by the $w_{\text{deadline}} \cdot I_{\text{ontime}}$ component of the reward function.

And the \textbf{communication requirement} ($\Omega_j$) is abstracted into a Communication Profile. This profile specifies the task's network sensitivity, including its communication topology (e.g., ring, all-to-all) and data volume. This high-level abstraction generates realistic communication penalties under various network conditions, directly informing the $P_{\text{comm}}$ penalty factor in our reward function and challenging the agent to make network-aware placements.

\subsection{Simulation of a Dynamic and Non-Stationary Environment}
A key differentiator of our platform is its ability to simulate a non-stationary environment, reflecting the dynamic nature of community networks. We model the network as a dynamic graph where inter-regional link behavior is driven by two distinct sources of non-stationarity. First, a \textbf{phased}, 24-hour model captures the systematic diurnal traffic patterns resulting from daily user activity cycles, with distinct periods like ``Afternoon Peak" or ``Overnight Batch" with specific workload characteristics. Complementing this predictable rhythm, a \textbf{probabilistic event injection mechanism} introduces stochastic disruptions, simulating unpredictable network outages or congestion bursts by temporarily and drastically reducing bandwidth on random links. Inter-region network latency is modeled using a base value derived from a static lookup table generated at initialization. To this base delay, we add minor stochastic fluctuations to simulate realistic network conditions.

\begin{table}[t]
\centering
\caption{State of the GPU Resource Pool}
\label{tab:case_study_gpus}
\begin{tabular}{llp{1.2cm}l}
\toprule
\textbf{GPU Group} & \textbf{Location} & \textbf{Trap} & \textbf{Notes} \\
\midrule
A100 1-2 (2)   & Asia-East  & Performance     & \makecell[l]{High compute,  far from data.} \\ 

A100 3-4 (2)   & US-East    & Reliability     & Co-located, low reliablity. \\

\textbf{4090 5-6 (2)}  & \textbf{US-East}      & \textbf{Optimal}  & \textbf{\makecell[l]{Low cost,  good performance.}} \\

\bottomrule
\end{tabular}
\end{table}

\subsection{Case Study: Dissecting Scheduling Decisions}
To illustrate the practical differences in scheduling logic, we present a case study for a communication-intensive task that requires four set of GPUs, with its dataset located in \texttt{US-East}. The state of the GPU resource pool is detailed in ~\autoref{tab:case_study_gpus}.

A \textbf{Greedy} scheduler aiming to maximize instantaneous throughput, and the \textbf{Random} and \textbf{Round-Robin} schedulers relying on stochasticity and fixed polling respectively, all exhibit suboptimal decision-making despite their distinct internal logics. Their fundamental flaw lies in their complete disregard for data locality and network topology. This inherently myopic design inevitably results in high data egress costs and significant communication latency.


In contrast, REACH demonstrates a learned understanding of the environment's trade-offs. Its policy has been trained to associate assignments far from the data source with a lower reward. Therefore, while both the \texttt{Asia East} and \texttt{US East} groups offer high compute capacity, REACH correctly identifies the \texttt{US East} group as the superior choice. This decision to co-locate computation with data simultaneously maximizes performance and minimizes operational costs, showcasing REACH's ability to achieve a globally optimal allocation where simpler strategies fail.

\begin{figure}[tbp]
    \centering
    \includegraphics[width=0.5\textwidth]{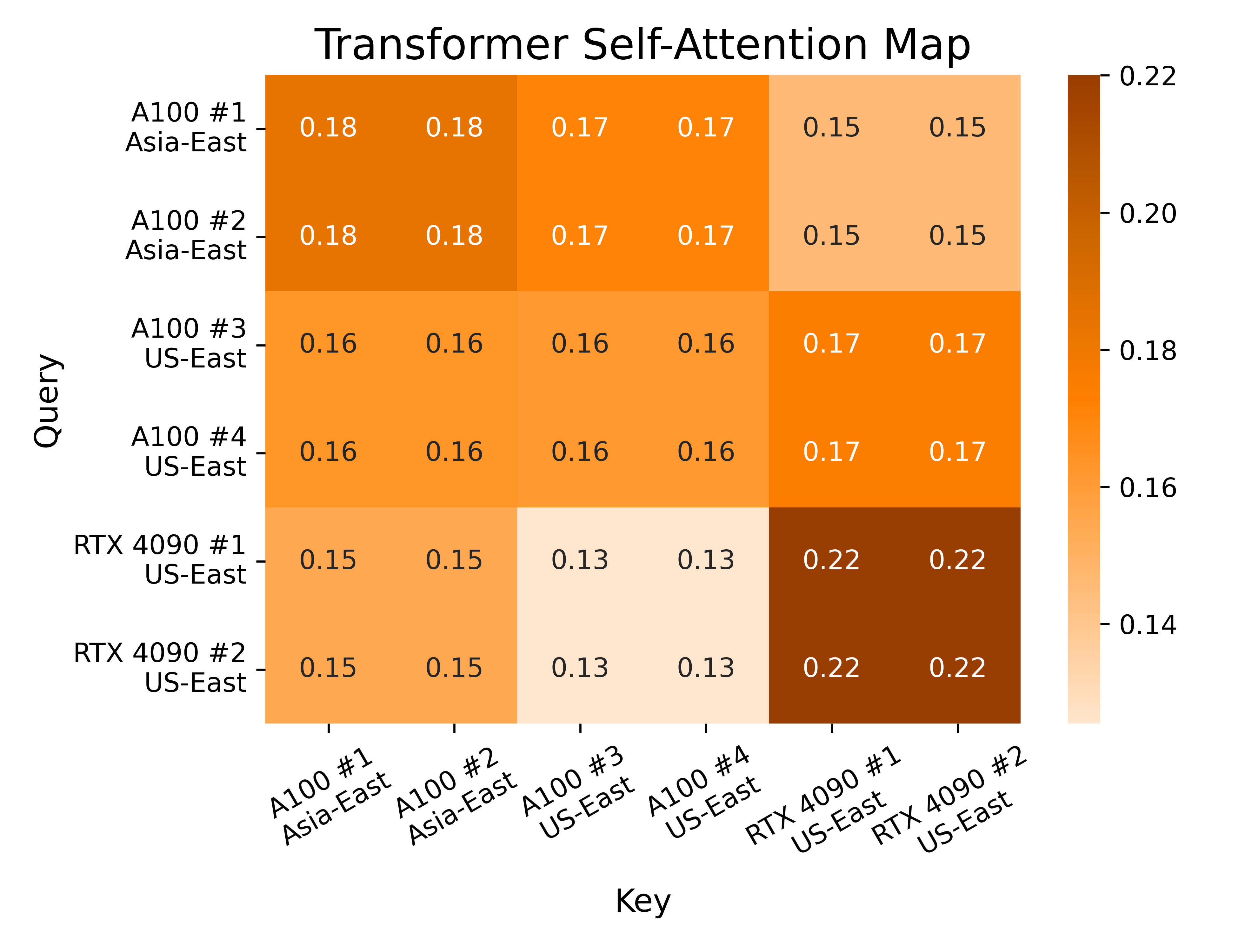}
    \caption{Averaged self-attention weights across all heads in REACH's Transformer. Higher weights indicate stronger influence when making assignment decisions.
}
    \label{fig:attention-map}
\end{figure}

As shown in~\autoref{fig:attention-map}, the averaged self-attention weights reveal how REACH evaluates the trade-offs among available GPUs. 
While the \textit{A100} groups offer the highest raw compute capability, the \texttt{Asia-East} A100s suffer from poor data locality due to their distance from the dataset, and the \texttt{US-East} A100s, though co-located, have low reliability. 
In contrast, the \textit{RTX 4090} nodes in \texttt{US-East} combine strong performance with the lowest cost and stable availability. 
REACH’s learned policy reflects this balance: despite the A100s have superior peak performance, the 4090 \texttt{US-East} nodes receive the highest attention scores.

The attention heatmap thus provides interpretability into the model's decision-making: REACH weighs global GPU capabilities while remaining sensitive to cost and latency trade-offs. This contrasts with heuristic strategies, which lack the capacity to perform such nuanced reasoning.



\section{Evaluation}


To evaluate the design and implementation of REACH, We structure the evaluation around three axes—performance, robustness/generalization, and scalability/architecture—detailing metrics, stress tests, and ablations for each.

\noindent \textbf{Core scheduling performance.} This involves not only measuring its overall \textit{efficiency} improvements against baseline strategies in terms of metrics like completion rate, deadline satisfaction, and GoodPut, but also conducting a fine-grained analysis of \textit{task-awareness}, examining how it intelligently adapts its scheduling decisions based on diverse task characteristics such as criticality, communication intensity.

\noindent \textbf{Robustness and Generalization.} We systematically test REACH's resilience under adverse environmental conditions, such as escalating GPU dropout rates and network congestion. Concurrently, we subject it to a variety of new workload patterns not seen during training to assess the generalization capabilities of its learned policy.

\noindent \textbf{Scalability and Architectural effectiveness.} We validate whether the REACH algorithm's performance can scale effectively by testing it in a large-scale, high-contention simulated environment. Furthermore, we conduct an architectural \textit{ablation study}, comparing it against a simplified MLP-based model to quantify the critical role of its Transformer core in capturing complex spatiotemporal dependencies and enhancing scheduling quality.

\begin{figure}[t]
\centering

\subfloat[\label{tab:agent_hyperparameters}]{
\begin{minipage}[t]{0.4\linewidth}
\vspace{0pt}
\scriptsize
\renewcommand{\arraystretch}{0.95}
\begin{tabular}{@{}p{2.4cm}p{1cm}@{}}
\hline
\textbf{Parameter} & \textbf{Value} \\
\hline
\multicolumn{2}{c}{\textit{PPO Learning Parameters}} \\
Learning Rate      & $3 \times 10^{-4}$ \\
Gamma ($\gamma$)   & 0.99 \\
Batch Size         & 32 \\
PPO Epochs         & 4 \\
PPO Clip Epsilon ($\epsilon$)   & 0.2 \\
Entropy Coefficient     & 0.01 \\
Value Loss Coefficient  & 0.5 \\
\hline
\multicolumn{2}{c}{\textit{Model Architecture}} \\
$d_{model}$           & 128 \\
$n_{head}$ & 4 \\
Encoder Layers & 2 \\
\hline
\end{tabular}
\end{minipage}
}
\hspace{0.1em}
\subfloat[\label{fig:training_loss}]{
\begin{minipage}[t]{0.47\linewidth}
\vspace{0pt}
\centering
\includegraphics[width=\linewidth]{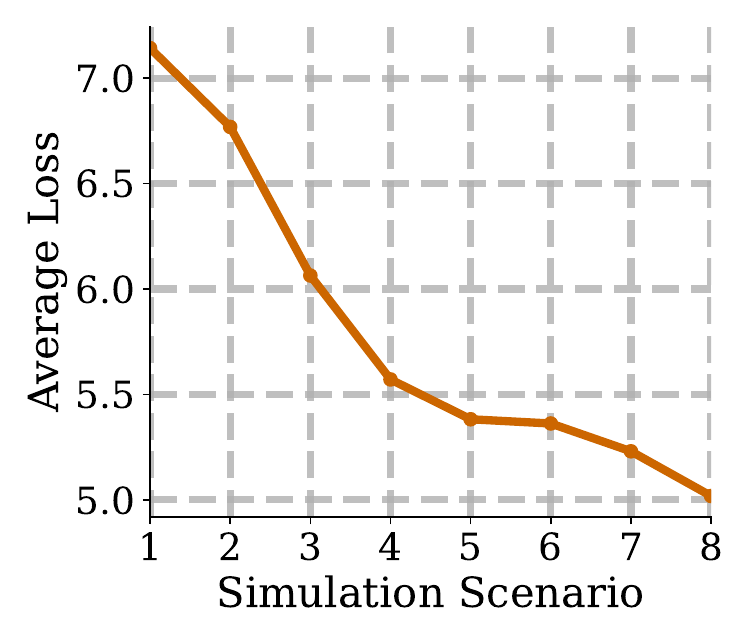}
\end{minipage}
}

\caption{REACH (a) agent configuration and (b) loss over simulation trainings.}
\label{fig:agent_config}
\end{figure}

\subsection{Experimental Setup}
We define two core experimental setups: \textbf{Small-scale} and \textbf{Large-scale}, simulating a spectrum of market conditions ranging from resource-constrained to resource-rich environments.

 All experiments utilize a consistent, phased workload generation mechanism designed to mimic the non-uniform arrival patterns of real-world tasks. This workload includes distinct phases such as "morning sessions," "afternoon peaks," and "overnight batch processing," each with a different mix of task types, computational demands, and priorities. All jobs are sampled from a predefined library of realistic task templates~\cite{bert_finetune,Qlora_finetune_technique,distillBert_multiGPU_finetune,stabilityai_inference,radford2023robust,rombach2022high,pugetsystems_blender_vram}.

\begin{figure*}[t]
  \centering
  \includegraphics[width=\linewidth]{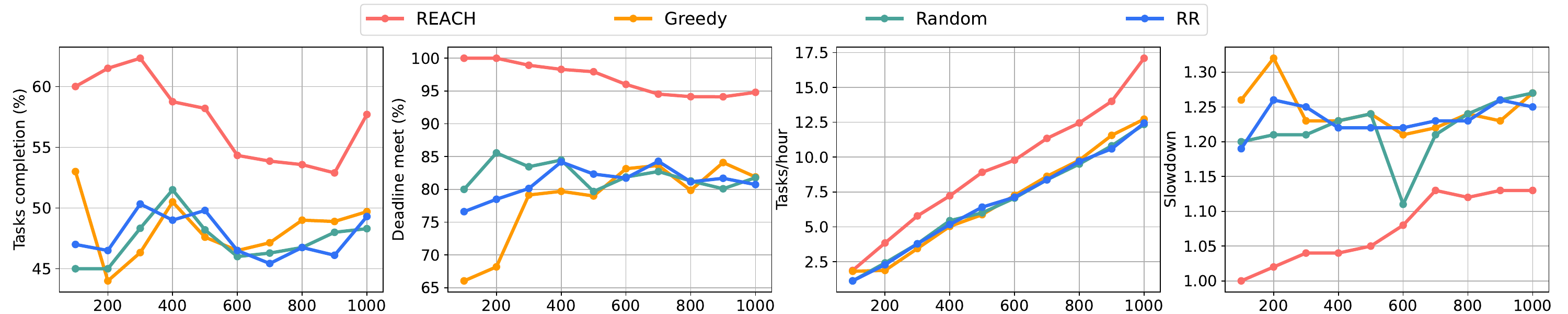}
  
  \parbox[t]{0.23\linewidth}{\centering \footnotesize \textbf{(a)} Completion Rate\\[-1pt]}
  \hfill
  \parbox[t]{0.23\linewidth}{\centering \footnotesize \textbf{(b)} Deadline Satisfaction\\[-1pt]}
  \hfill
  \parbox[t]{0.23\linewidth}{\centering \footnotesize \textbf{(c)} GoodPut\\[-1pt]}
  \hfill
  \parbox[t]{0.23\linewidth}{\centering \footnotesize \textbf{(d)} Job Slowdown\\[-1pt]}

  \caption{REACH consistently enhances platform-wide performance across scheduling metrics.}
  \label{fig:scheduling_metrics_row}
\end{figure*}


The key hyperparameters used for training our REACH agent are detailed in ~\autoref{tab:agent_hyperparameters}. During training, the REACH agent’s loss converged smoothly across simulation scenarios, as shown in \autoref{fig:training_loss}, indicating stable policy optimization. This confirms that the learning process was stable before evaluating scheduling performance.

\subsection{Baseline and Metrics}
To the best of our knowledge, no mature, widely-accepted scheduling frameworks for community GPU platforms that can serve as strong baselines. Prior studies in centralized GPU clusters typically assume stable, homogeneous, and well-connected resources, making them unsuitable for the volatility, heterogeneity, and dynamic community environments. 

Given the absence of specialized baselines for this setting, we select three representative strategies—Greedy, Random, and Round-Robin—that capture distinct and fundamental scheduling philosophies: greedy optimization, stochastic allocation, and load balancing. These strategies provide a clear and interpretable reference point, enabling us to evaluate the advantages of our proposed method within a transparent and reproducible comparative context. \texttt{\textbf{Greedy Strategy}:} This strategy always prioritizes selecting the combination of GPUs with the highest compute power for a task. Its core logic is that the most powerful hardware should yield the shortest theoretical computation time. \texttt{\textbf{Random Strategy}:} This strategy selects the required number of GPUs for a task completely at random from all available resources that meet the basic requirements such as memory and online status. \texttt{\textbf{Round Robin Strategy}:} This strategy maintains a globally consistent list of GPUs and allocates resources sequentially using an internal pointer. Its design objective is to achieve long-term load balancing. 

The common limitation of these baselines is their reliance on static, single-dimensional rules, rendering them unable to adapt to the dynamic and multi-faceted environment our platform simulates. This inherent myopia provides a clear comparative context for demonstrating the advanced, adaptive capabilities of the REACH algorithm.

To quantify scheduler performance, we adopt a suite of metrics. For evaluating overall efficiency and robustness, we focus on four primary system-level indicators:
    (a) \textbf{\texttt{Completion Rate}:} The percentage of submitted tasks that complete successfully without errors.
    (b) \textbf{\texttt{Deadline Satisfaction}:} Among all successfully completed tasks, the fraction that finished within their deadline.
    (c) \textbf{\texttt{GoodPut}:} The number of successfully completed tasks per unit of time, reflecting the platform's effective throughput.
    (d) \textbf{\texttt{Job Slowdown}:} The ratio between a task's actual turnaround time and its ideal execution time, measuring user-perceived responsiveness.

For more specialized analyses, such as task-type awareness, we utilize targeted metrics and visualizations like turnaround time CDFs and resource allocation distributions, which will be introduced in their respective sections.



\subsection{Core Scheduling Performance}
To assess REACH's core performance, we begin by evaluating its overall scheduling efficiency. This initial analysis focuses on four system-level metrics: task success rate, deadline satisfaction rate, GoodPut, and mean job slowdown. Together, these metrics provide a holistic view of the system's throughput, responsiveness, and service quality under various loads. Following this, we will conduct a more fine-grained task-aware analysis to understand how REACH achieves these results by catering to the specific needs of different job types. 

\subsubsection{Overall Scheduling Efficiency}

\noindent \textbf{Highest task completion across all loads, with a clear edge under light workloads.}  
As shown in \autoref{fig:scheduling_metrics_row}, under a 100-task scenario, REACH completes 59.8\% of submitted jobs, while the next-best strategy, Greedy, achieves only 53\%—a gap of over 6 percentage points. As the task load increases to 1000, all strategies experience performance degradation, but REACH maintains the lead, completing 57.7\% of tasks compared to 49.2\% for Greedy, which suffers from high deadline violation and failure rates. These results are consistent with higher completion rate across all workloads.

\noindent \textbf{Higher deadline satisfaction and consistent QoS target achievement.}  
REACH consistently outperforms all baseline strategies in deadline satisfaction across varying system loads. While all strategies show a slight decline as task load increases, REACH remains above 95\% throughout, compared to others fluctuating between 66\% and 85\%.  Even under the heaviest load, REACH maintains a high rate of 95.1\%, while the baselines fall below 83\%.  Among all workloads, REACH outperforms baselines about 10\% to 25\%.

\begin{figure}[t]
    \begin{minipage}[htbp]{0.49\linewidth}
    \centering
        \includegraphics[width=\linewidth]{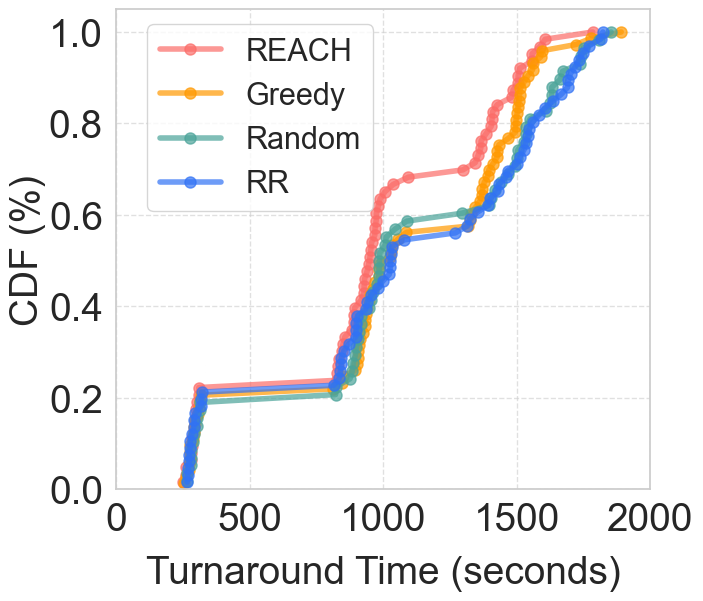}
    \caption{Turnaround time distribution for critical tasks. 
  }\label{fig:short_jobs}
    \end{minipage}
    \hfill
    \begin{minipage}[htbp]{0.49\linewidth}
    \centering
        \includegraphics[width=\linewidth]{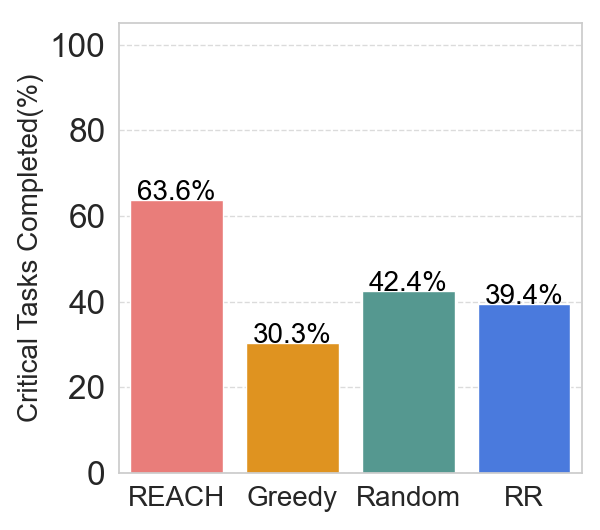}
    \caption{Critical Task Completion Rates comparison.}\label{fig:critical_tasks_p1_summary}
    \end{minipage}
\end{figure}

\noindent \textbf{Leading GoodPut and reduced computation waste.}  
REACH consistently delivers the highest GoodPut across all load levels, with the gap widening as the workload increase-about 30\% higher than the nearest baseline above 400 tasks workload. At 1000 tasks, REACH achieves 16.8 tasks/hour, significantly ahead of all baselines. 

\noindent \textbf{Minimal slowdown, even under heavy load.}
REACH consistently achieves the lowest slowdown across all load levels, with values rising modestly from 1.02 to 1.14 as the system becomes busier. In contrast, all baselines stay above 1.2 and exhibit larger fluctuations. The slowdown gap indicates that REACH keeps job delays more uniform across job sizes, with smaller jobs avoiding the disproportionate slowdowns observed in all baselines under high contention.

\subsubsection{Task-Type Aware Scheduling}

To explain why REACH achieves such strong overall results, we now analyze its behavior at the task level. To understand the source of these gains, we next examine REACH's performance on three distinct task types.



\noindent \textbf{Consistently low turnaround times for critical tasks.} ~\autoref{fig:short_jobs} plots the turnaround time CDF of critical tasks. REACH outperforms all baselines by completing a larger fraction of jobs within tighter latency bounds. Over 90\% of REACH-scheduled critical tasks finish under 1000 seconds, while others show heavier tails. This indicates that REACH recognizes task urgency and reduces queueing delays without sacrificing overall fairness. Furthermore, REACH's superiority is not limited to speed; it also achieves a significantly higher success rate. \autoref{fig:critical_tasks_p1_summary} shows that REACH completes 63.6\% of critical tasks, more than doubling the rate of the Greedy (30.3\%) and substantially outperforming others.

\begin{figure}[t]
    \centering
    \begin{subfigure}[b]{0.48\linewidth}
        \centering
        \includegraphics[width=\linewidth]{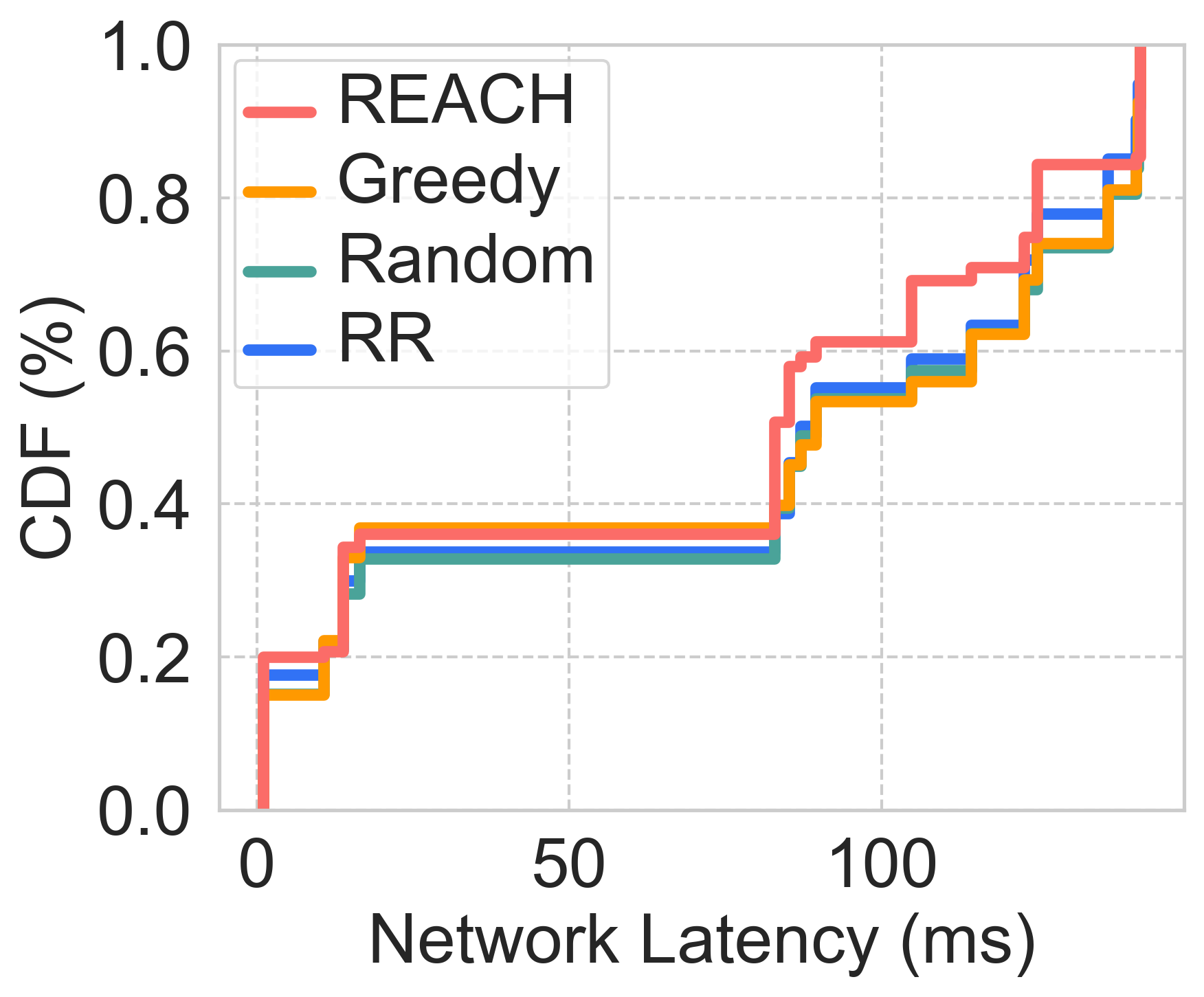}
        \caption{}
        \label{fig:network_latency_cdf}
    \end{subfigure}
    \hfill
    \begin{subfigure}[b]{0.48\linewidth}
        \centering
        \includegraphics[width=\linewidth]{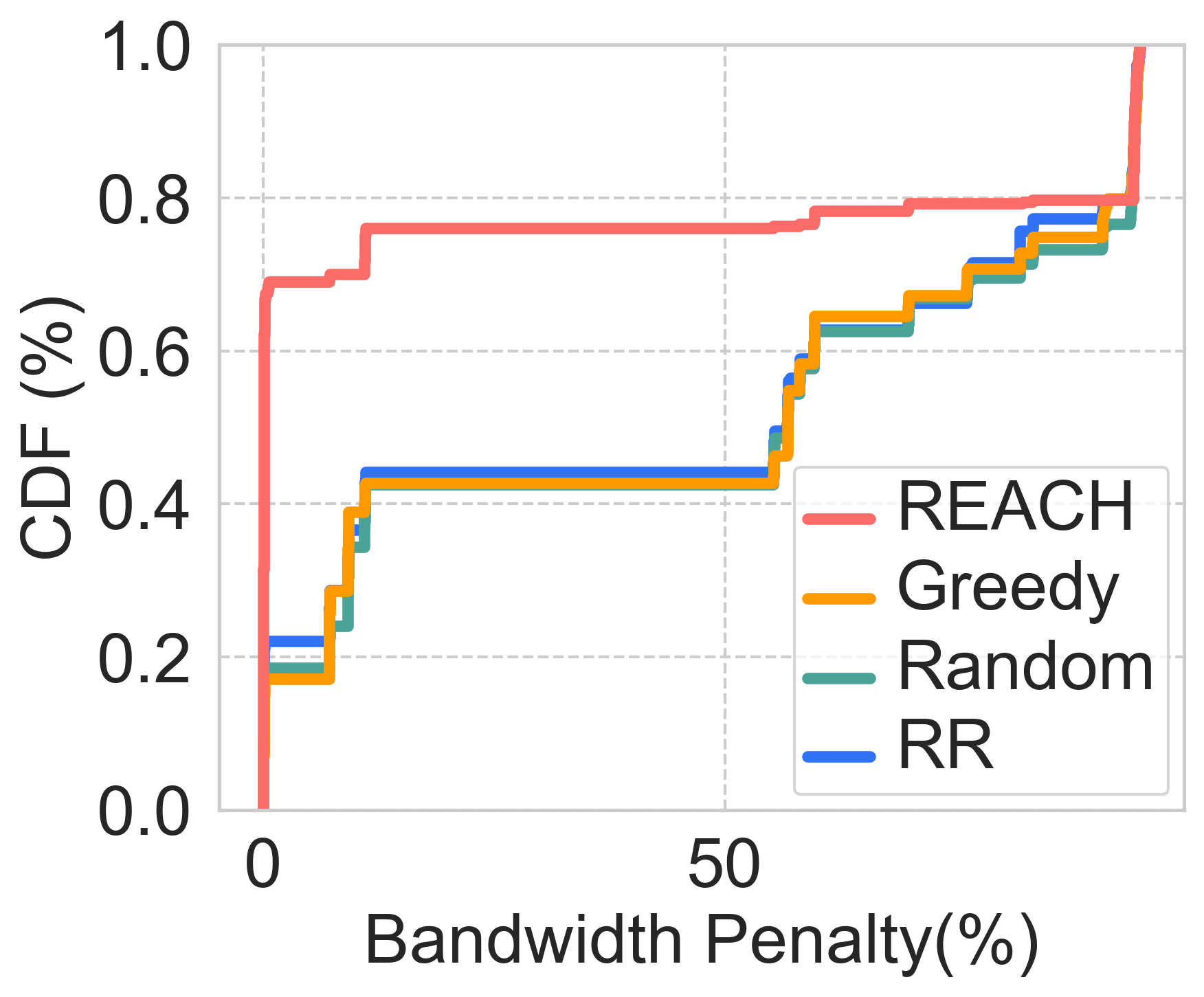}
        \caption{}
        \label{fig:bandwidth_penalty}
    \end{subfigure}
    \caption{Latency and bandwidth penalty analysis of different scheduling strategies.}
    \label{fig:combined-network-bandwidth}
\end{figure}

\noindent \textbf{Topology-aware placement reduces communication penalties.} Communication-intensive tasks are highly sensitive to GPU placement due to large internode traffic. As shown in ~\autoref{fig:network_latency_cdf} and ~\autoref{fig:bandwidth_penalty}, while the latency distributions for all strategies are largely similar, REACH shows a significant advantage in mitigating bandwidth penalties. For REACH, over 80\% of tasks incur lower than 5\% bandwidth penalty, indicating exceptionally effective co-location of communicating jobs. In stark opposition, baseline methods show a wide spread of penalties, with a large fraction of their tasks suffering penalties greater than 60\%. This allocation pattern reduces communication overhead and is associated with higher task success rates under heavy load.

\begin{figure}[t]
  \centering
  \includegraphics[width=0.95\linewidth]{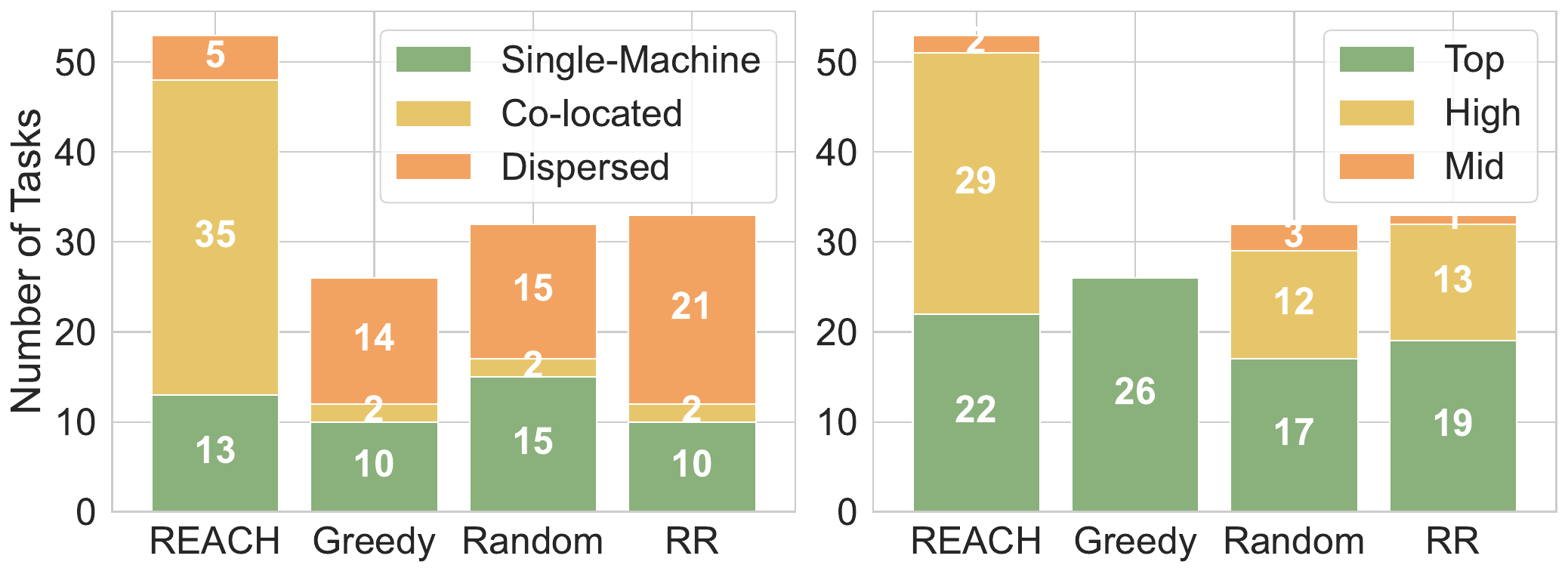}
  \parbox[t]{0.4\linewidth}{\footnotesize \textbf{(a)} GPU affinity distribution for large-scale tasks.}
  \hspace{4mm}
  \parbox[t]{0.4\linewidth}{\centering \footnotesize \textbf{(b)} GPU ability distribution. }
  \caption{
  A comparison of resource allocation for large-scale tasks across different scheduling strategies.
  }
  \label{fig:GPU affinity and ability distribution}
\end{figure}

\noindent \textbf{Locality-aware, flexible dispatch for large-scale jobs.} 
For large-scale workloads like model-parallel training, inter-GPU communication is not merely an overhead but part of the computational critical path. Each training step requires synchronous data exchange between GPUs holding different model shards. Placing these shards across a slow network drastically increases synchronization overhead, effectively stalling the entire computation and nullifying the benefit of having more GPUs.
Therefore, a scheduler's ability to assemble a set of GPUs with high-bandwidth interconnects is paramount. The allocation data in \autoref{fig:GPU affinity and ability distribution} reveals REACH's intelligent approach to this challenge. While it places slightly fewer jobs on a single machine—the ideal but often scarce option—it excels at maximizing co-located dispatches to ensure a high-speed critical path. In contrast, baseline methods that ignore this requirement often create unusable resource bundles. This demonstrates REACH's advantage in consolidating not just raw compute power, but usable, high-throughput compute clusters for the most demanding workloads.

\subsection{Robustness and Generalization}

A critical requirement for any real-world scheduler is the ability to perform reliably under unpredictable conditions. To validate this capability in REACH, we conduct two sets of stress tests. First, we evaluate its resilience by subjecting it to deteriorating environmental conditions. Second, we test its generalization by exposing it to workload patterns fundamentally different from its training data.

\subsubsection{Resilience to Environmental Instability} 
To quantify its resilience, we systematically increase the frequency of adverse events and observe the impact on key performance metrics.

\begin{figure}[t]
  \centering
  \includegraphics[width=\linewidth]{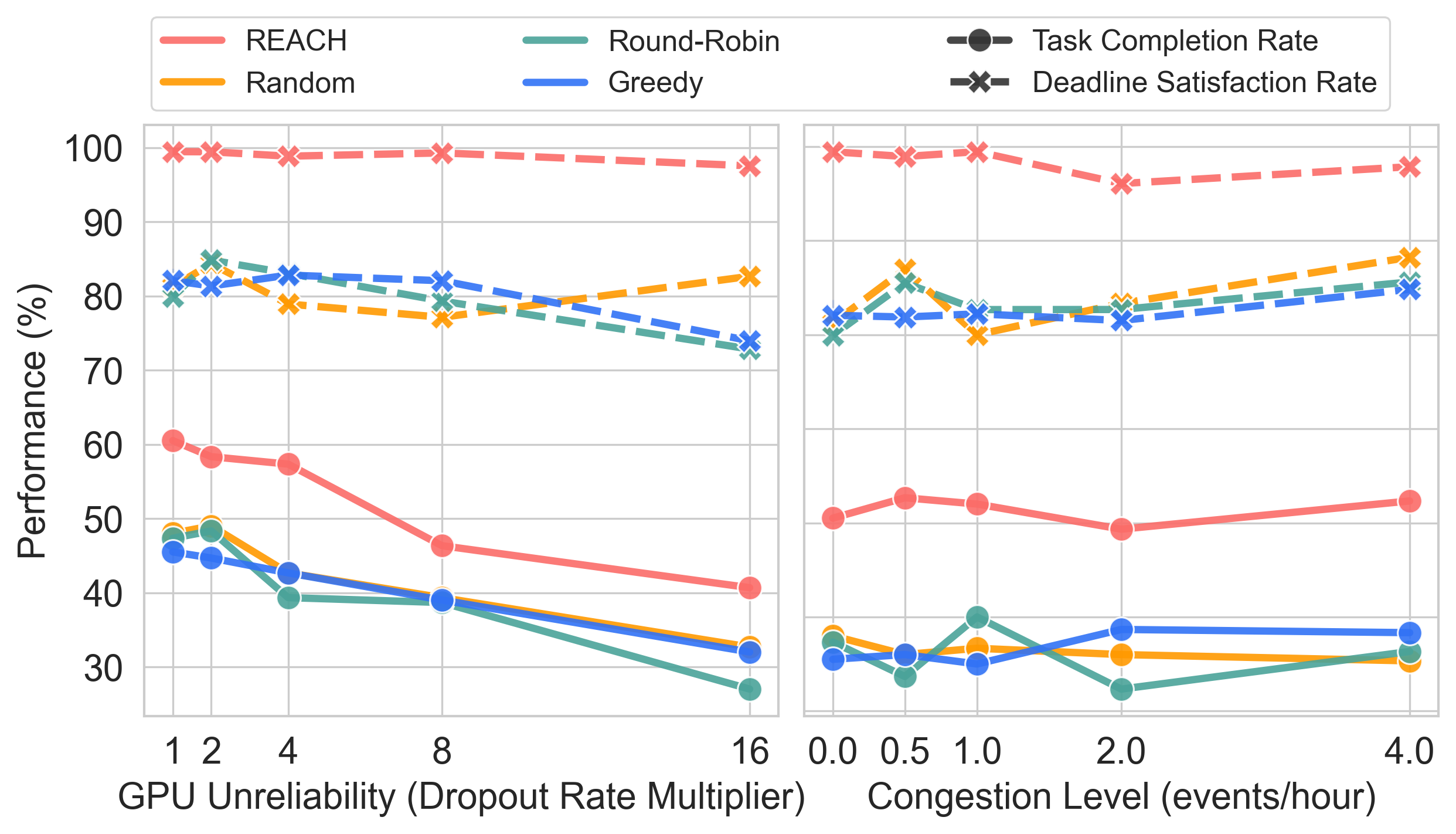}
  \caption{Performance under deteriorating conditions: (a) increasing GPU unreliability and (b) escalating network congestion.}
  \label{fig:robustness_analysis}
\end{figure}

 \noindent \textbf{Superior resilience against increasing GPU unreliability.}
As shown in \autoref{fig:robustness_analysis}, while the task completion rate for all strategies declines with a GPU dropout rate multiplier increasing from 1$\times$ to 16$\times$, REACH consistently maintains a advantage. Its completion rate stays around 40\% even at the highest unreliability, compared to under 30\% for baseline strategies. More strikingly, its deadline satisfaction rate shows exceptional stability, remaining over 95\% across all tested levels, which is in stark contrast to the baseline strategies whose satisfaction rates show a more pronounced decline.

\noindent \textbf{Stable performance under escalating network congestion.}
\autoref{fig:robustness_analysis} shows that REACH's task completion rate holds steady at approximately 50\%, largely unaffected by the increase in network congestion events, while baseline strategies perform at a much lower and more volatile level. The baseline strategy exhibits fluctuating performance across congestion levels, with its satisfaction rate ranging from 25\% to 40\%, while showing intermittent variations at intermediate congestion points.

\subsubsection{Generalization to Unseen Workload Patterns} 
\begin{figure*}[t]
    \centering
    \includegraphics[width=\textwidth]{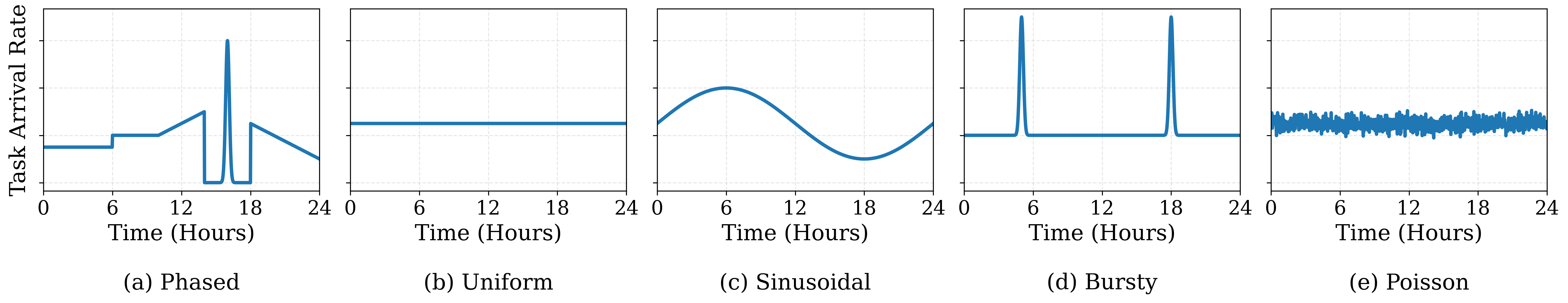}
    \caption{Five workload models over a 24-hour cycle.} 
    \label{fig:workload-profiles}
\end{figure*}

To test the model's generalization beyond its training data, we evaluated it on four distinct unstructured workloads designed to differ significantly from the phased training environment. \autoref{fig:workload-profiles} visualizes these test scenarios (from b to e) alongside the original phased training workload \textbf{(a)}. The test workloads are defined as follows: \textbf{(b) Uniform Workload}: A completely patternless environment by uniformly sampling all task properties (arrival time, type, priority) across their entire ranges. It serves as a fundamental generalization baseline. \textbf{(c) Sinusoidal Workload}: A smooth, continuous load cycle where the task arrival rate follows a sinusoidal function, testing adaptation to non-linear, gradual dynamics. \textbf{(d) Bursty Workload}: Characterized by high-intensity task bursts concentrated in short time windows over a low-level background load, simulating high-volatility scenarios and sudden system pressure. \textbf{(e) Poisson Workload}: Treats task arrivals as a memoryless Poisson process, where inter-arrival times are drawn from an exponential distribution.

\begin{figure}[t]
    \centering
    \includegraphics[width=\columnwidth]{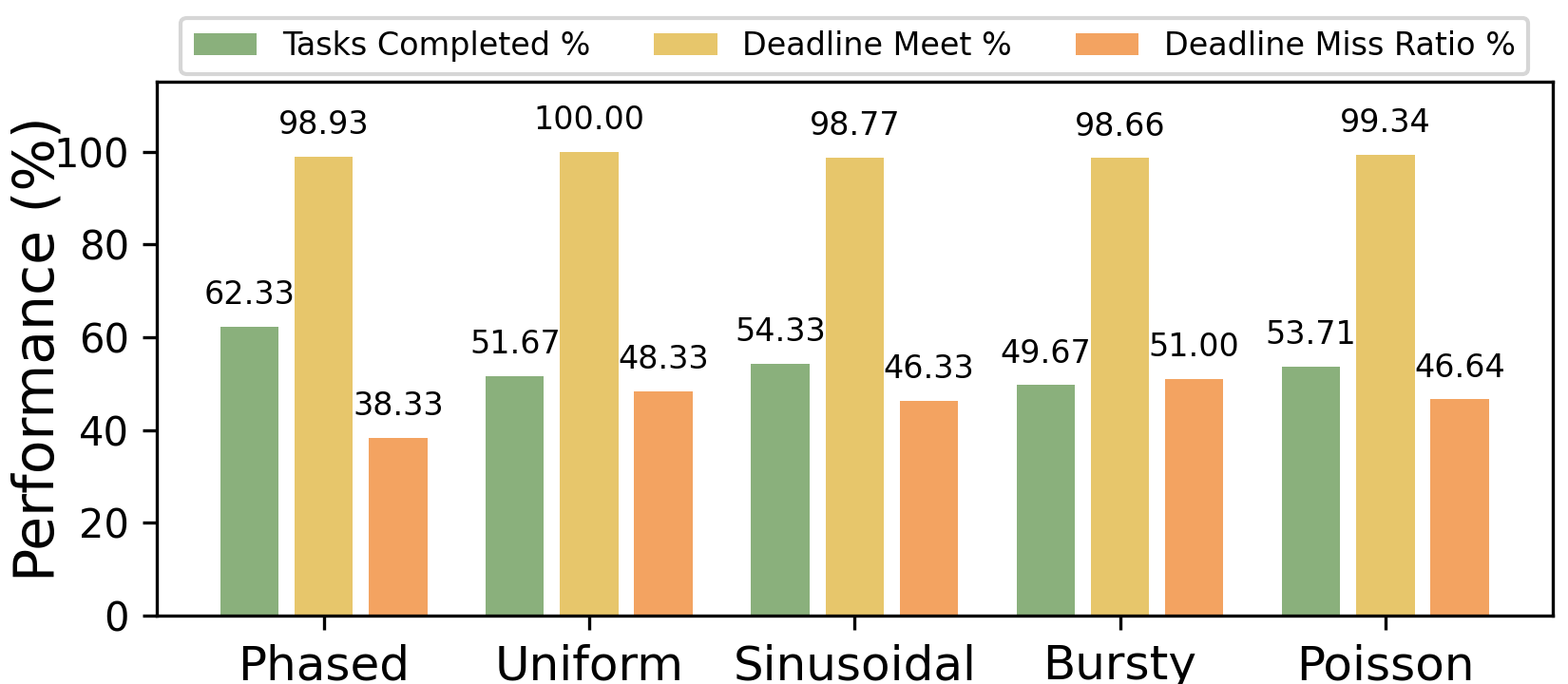}
    \caption{Performance of REACH across various workload patterns and untrained traffic models.}
    \label{fig:workload-performance}
\end{figure}

\noindent \textbf{Stable task completion across diverse traffic patterns.} While the percentage of tasks completed naturally varies with workload characteristics, REACH shows remarkable consistency. The completion rate remains within a stable range, from 49.67\% under the Bursty workload to 62.33\% in the Phased scenario, with no drastic performance collapses. This indicates that the model's learned scheduling logic is not overfitted to a specific traffic type but is robust enough to manage a variety of real-world task arrival scenarios.

\noindent \textbf{Near perfect deadline satisfaction in completed tasks across unseen workloads.} As shown in \autoref{fig:workload-performance}, the Deadline Met Rate for REACH remains consistently stay above 98.5\% across all four challenging and distinct workload patterns—Uniform, Sinusoidal, Bursty, and Poisson. However, the overall Deadline Miss Ratio ranges from 38.33\% to 51.00\%, reflecting that a considerable fraction of tasks still fail to meet deadlines under heavy contention, primarily due to task incompletion rather than late completion.

\subsection{Scalability and Architectural Analysis}
To complete our evaluation, we push REACH to its limits and dissect the architectural choices underpinning its performance. This final analysis validates two critical aspects: first, we assess its scalability in a demanding, high-contention environment. Second, through an architectural ablation study, we isolate and quantify the contribution of the Transformer core to understand the source of its intelligent decision-making.

\subsubsection{Scalability in Large-Scale Environments} 
To test REACH's scalability under high-contention conditions, we simulated a large-scale environment consisting of 1,000 heterogeneous GPUs and a workload of 5,000 incoming tasks. As illustrated in \autoref{fig:large scale network lab}, the results demonstrate the comprehensive superiority of the REACH strategy, which significantly envelops all baselines across every evaluated dimension.

\noindent \textbf{QoS and cost-efficiency under high contention.}
In ~\autoref{fig:radar_chart}, REACH attains the largest radar chart area among all methods, with the highest observed values in deadline satisfaction and cost-efficiency when network scale and task volume are increased. These results suggest that the learned policy allocates resources in a way that maintains service-level compliance while reducing loss relative to baselines.

\noindent \textbf{Throughput and responsiveness.}
REACH also achieves higher GoodPut, completion rate, and responsiveness (1/slowdown) than the other three strategies, as reflected by its extended axis lengths in the radar plot. In comparison, baseline metrics are concentrated near the chart center, indicating lower task throughput and slower response under the same conditions.

\noindent \subsubsection{Architectural Ablation Study} To isolate and understand the impact of our model's internal design, we conduct an ablation study comparing the full Transformer model to a version where its core is replaced with a simpler MLP architecture. This experiment quantifies how much of the observed performance gain stems from the Transformer's ability to capture complex spatiotemporal dependencies, rather than other heuristics.

\begin{figure}[t]
    \begin{minipage}[htbp]{0.49\linewidth}
    \centering
        \includegraphics[width=\linewidth]{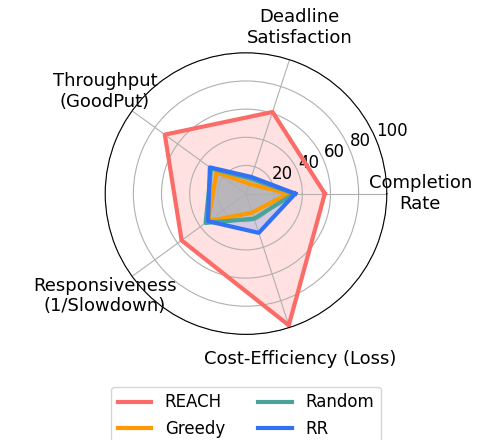}
    \caption{The REACH strategy performs excellently in large-scale network tests.}\label{fig:large scale network lab}
    \end{minipage}
    \hfill
    \begin{minipage}[htbp]{0.49\linewidth}
    \centering
        \includegraphics[width=\linewidth]{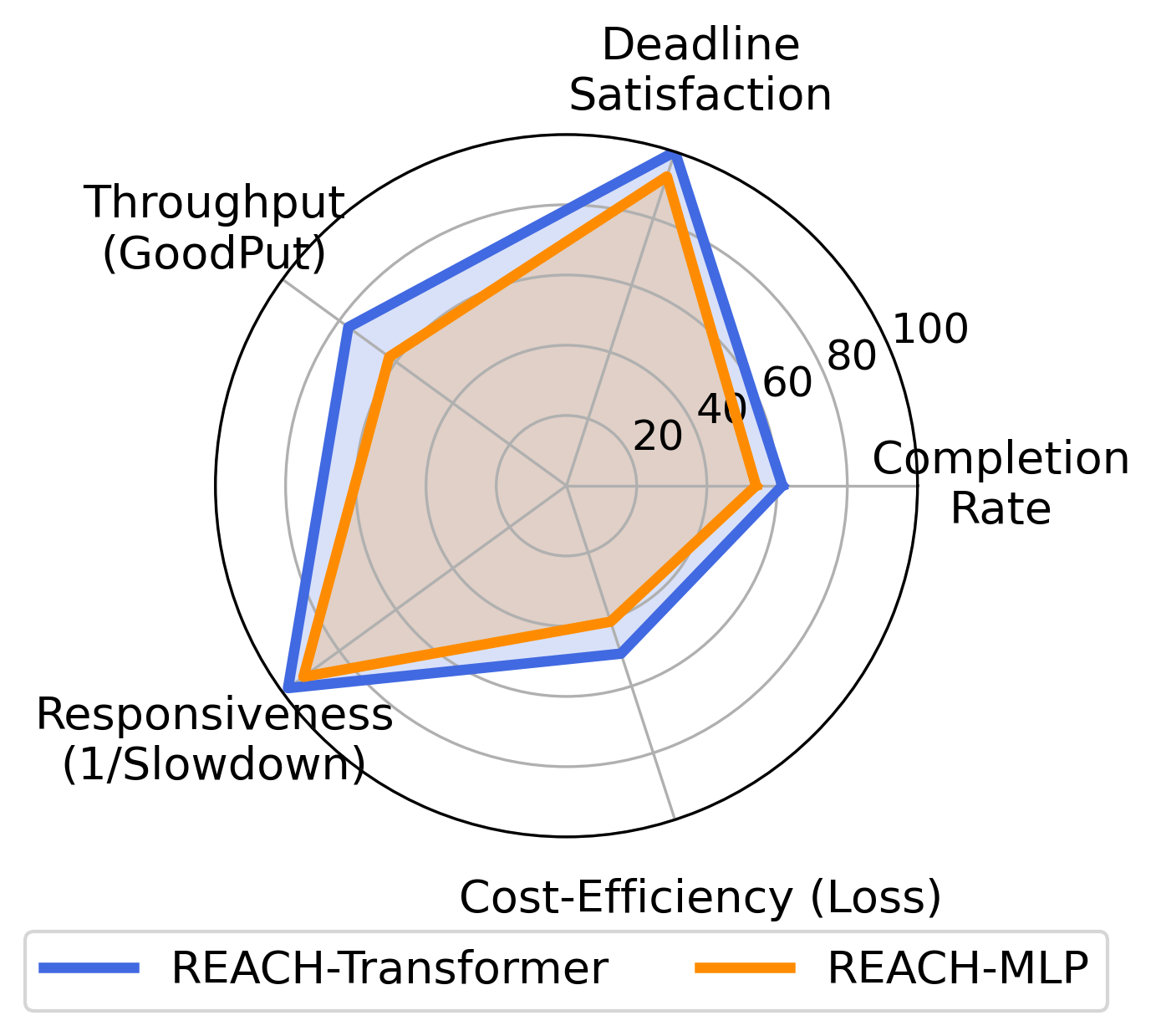}
    \caption{Performance comparison of REACH base on Transformer and MLP layers. 
  }\label{fig:radar_chart}
    \end{minipage}
    \vspace{-0.20in}
\end{figure}

\noindent \textbf{Transformer core as key to superior quality-of-service.}
The results, presented in \autoref{fig:radar_chart}, confirm the vital role of our architectural choice. The Transformer model demonstrates a clear and significant performance advantage over the MLP baseline, particularly in the key quality-focused metrics of Deadline Satisfaction, GoodPut, and Responsiveness. While both architectures exhibit comparable cost-efficiency and raw completion rates, the Transformer’s ability to excel in these crucial areas highlights its effectiveness in making more intelligent, context-aware scheduling decisions, thereby validating its integral role within the REACH framework.




\section{Discussion}

This section provides a high-level summary of the experimental findings and offers deeper insights into the mechanisms behind REACH's performance, along with the broader implications of this research.

\subsection{Interpretation of Key Results}

Across a wide range of experiments, REACH consistently outperforms traditional static scheduling baselines in terms of all metrics. The performance gap becomes more pronounced under challenging conditions, such as fluctuating workloads, dynamic network delays, and unstable GPU nodes. These results suggest that REACH not only learns optimal resource-task mappings but also possesses a strong ability to adapt to real-time platform dynamics.

The ablation study further underscores the critical role of the Transformer architecture. When the scheduling policy replaces the Transformer with a simple MLP, performance drops significantly, especially under conditions of resource contention or node volatility. This highlights the value of the Transformer's global modeling capability, enabling REACH to simultaneously perceive complex inter-GPU relationships—such as topology proximity, reliability history, and load distribution—and make collaborative, context-aware scheduling decisions.

\subsection{Mechanisms Behind Generalization and Robustness}

REACH maintains performance under unseen workloads and unexpected environmental changes. This generalization does not stem from memorizing specific traffic patterns, but rather from its learned scheduling behaviors that are inherently context-aware. By leveraging information about task types, prior dispatch outcomes, and node-level statistics, REACH adapts its decisions dynamically and avoids over-reliance on superficially favorable but historically unreliable nodes.

This explains REACH's robustness in stress tests. When network congestion or node failure rates increase, REACH effectively circumvents problematic nodes and adjusts its policy accordingly, leading to improved task reliability and QoS guarantees.

\subsection{Implications of the Findings}

The findings support an important insight: in highly heterogeneous and dynamic community GPU environments, intelligent scheduling via deep reinforcement learning can outperform static heuristics while maintaining strong generalization and fault tolerance. This not only demonstrates the practical viability of such platforms for critical tasks, but also provides valuable guidance for broader distributed computing domains—such as edge computing and autonomous compute fabrics—where resource heterogeneity and unreliability are common challenges.

\section{Limitations and Future Work}

Despite the promising results, this work presents several limitations that open up opportunities for future research and practical enhancements.

\noindent \textbf{Beyond Discrete-Event Simulation.}  
Our platform is based on discrete-event modeling, which captures high-level scheduling dynamics but omits fine-grained network behavior. Real-world systems involve complex protocol-level interactions, contention, and routing behaviors that are better captured by network simulators such as NS-3~\cite{henderson2008ns3} or MiniNet~\cite{lott2010mininet,yan2015vtmininet}. Future work should integrate REACH with these simulation platforms—or better, conduct real-system deployment and testing—to validate scheduling performance under actual latency, congestion, and failure scenarios.

\noindent \textbf{Security and Trust in Open Networks.}
The current framework assumes honest behavior and independent failures across nodes. In open community networks, however, malicious actors may deliberately misreport their state or exhibit coordinated dropouts. A future direction is to integrate trust modeling or adversarial reinforcement learning, enabling REACH to identify and mitigate unreliable nodes, thus improving robustness and long-term platform stability.

\noindent \textbf{Handling Unforeseen and Burst Events.}
While REACH exhibits robustness under predefined stress tests, it lacks explicit mechanisms for sudden workload bursts, network blackouts, or coordinated node failures. Future work could incorporate event-driven adaptation, such as anomaly detection, predictive buffering, or fallback heuristics, to further improve the system’s responsiveness and QoS under unforeseen disruptions.

\section{Conclusion}

In this paper, we proposed \textbf{REACH}, a reinforcement learning-based scheduling framework tailored for community GPU platforms characterized by heterogeneity, volatility, and uncertainty. By incorporating Transformer-based global modeling and a context-aware policy network, REACH effectively learns to allocate GPU resources in a dynamic environment, outperforming traditional rule-based and static scheduling strategies across multiple metrics.

Extensive experiments demonstrate that REACH not only achieves high task success rates and deadline satisfaction under normal conditions, but also exhibits strong robustness in the face of network congestion and node failures. Furthermore, ablation studies confirm the critical role of the Transformer architecture in enabling REACH to reason over complex multi-node interactions.

Overall, this work highlights the potential of intelligent scheduling to transform community GPU platforms from opportunistic compute pools into reliable, QoS-aware infrastructure. We believe REACH paves the way for more adaptive and trustworthy resource management in broader distributed AI systems.

\bibliographystyle{plain}
\renewcommand{\bibfont}{\footnotesize}
\bibliography{references}

\end{document}